\documentclass[a4paper,11pt]{article}

\usepackage{geometry}
\usepackage{bm}
\usepackage{amsmath,amssymb}
\usepackage{graphicx}
\usepackage[noend]{algpseudocode}
\usepackage{algorithmicx,algorithm}
\usepackage[comma,authoryear]{natbib} 
\usepackage[labelfont=bf]{caption}
\usepackage{float}
\usepackage{multicol,lipsum}
\usepackage[switch,columnwise]{lineno}

\usepackage[T1]{fontenc}
\usepackage[utf8]{inputenc}
\usepackage{authblk}
\usepackage[colorlinks,linkcolor=blue,citecolor=blue,urlcolor=blue]{hyperref}
\usepackage[title]{appendix}
\usepackage[]{pifont}
\usepackage{lettrine}
\usepackage{hyperref}
\usepackage{setspace}

\graphicspath{{figures/}} % Location of the graphics files

\numberwithin{equation}{section}

\title{\textbf{{\LARGE Evaluating the evolution and inter-individual variability of infant functional module development from 0 to 5 years old}}}

\author[a]{\small Lingbin Bian }
\author[a]{\small Nizhuan Wang } 
\author[a]{\small Yuanning Li } 
\author [d,e,f,g]{\small Adeel Razi } 
\author[a]{\small Qian Wang }
\author[a,c]{\small Han Zhang }
\author[a,b,c,*]{\small Dinggang Shen }
\author[ ]{\small the UNC/UMN Baby Connectome Project Consortium }

\affil[a]{School of Biomedical Engineering \& State Key Laboratory of Advanced Medical Materials and Devices, ShanghaiTech University, Shanghai 201210, China}
\affil[b]{Shanghai United Imaging Intelligence Co., Ltd., Shanghai 200230, China}
\affil[c]{Shanghai Clinical Research and Trial Center, Shanghai 201210, China}
\affil[d]{Turner Institute for Brain and Mental Health, School of Psychological Sciences, Monash University, Australia}
\affil[e]{Monash Biomedical Imaging, Monash University, Australia}
\affil[f]{Wellcome Centre for Human Neuroimaging, University College London, United Kingdom}
\affil[g]{CIFAR Azrieli Global Scholars Program, CIFAR, Canada}

\affil[*]{\small Corresponding author: Dinggang Shen (Dinggang.Shen@gmail.com)}
\date{} % discard date

\begin{document}
\renewcommand{\figurename}{\textbf{Fig.}} % 
\maketitle
%\noindent {\footnotesize \textsuperscript{*} indicates corresponding author: lingbin.bian@monash.edu; \textsuperscript{\S} indicates joint senior author.}\\

   % \lipsum[1]

%\begin{center}
%\linenumbers
\renewcommand{\baselinestretch}{1.5}
\begin{abstract}
\setlength{\parindent}{0pt} \setlength{\parskip}{1.5ex plus 0.5ex
minus 0.2ex} %\noindent

The segregation and integration of infant brain networks undergo tremendous changes due to the rapid development of brain function and organization. Traditional methods for estimating brain modularity usually rely on group-averaged functional connectivity (FC), often overlooking individual variability. To address this, we introduce a novel approach utilizing Bayesian modeling to analyze the dynamic development of functional modules in infants over time. This method retains inter-individual variability and, in comparison to conventional group averaging techniques, more effectively detects modules, taking into account the stationarity of module evolution.  Furthermore, we explore gender differences in module development under awake and sleep conditions by assessing modular similarities. Our results show that female infants demonstrate more distinct modular structures between these two conditions, possibly implying relative quiet and restful sleep compared with male infants.

\end{abstract}

\section{Introduction}
%\lettrine[lines=3]{\textbf{I}}{ \textbf{ntroduction}} 

Magnetic resonance imaging (MRI) provides a non-invasive means to assess both the structure and function of the infant brain, serving as a vital tool in understanding the mechanisms of infant brain development. For instance, the Baby Connectome Project (BCP) released cross-sectional and longitudinal high-resolution multimodal MRI data from 500 typically developing subjects from birth to 5 years old \citep{Howell2019}. The Developing Human Connectome Project (dHCP), published MRI data from 1500 subjects ranging from 20 to 44 weeks of gestation \citep{Eyre2021}. These large-scale datasets contribute significantly to expanding our knowledge of early brain development and provide crucial insights into the origins and abnormal developmental trajectories of neurodevelopmental disorders such as autism, schizophrenia, and attention deficit hyperactivity disorder (ADHD).

Spontaneous neural activity is believed to be related to the intrinsic functional organization of the human brain \citep{Fox2007}. Functional MRI (fMRI) is one of the most common methods for capturing brain function. It measures the blood oxygen level dependent (BOLD) signal, providing an indirect in vivo measurement of neural activity \citep{Ogawa1990}. Early brain development in infancy is challenging to grasp, especially the patterns of how functional connectivity (FC) which quantifies the statistical interdependence among regional time series derived from the BOLD signals emerges and develops in the initial years after birth. Task-related fMRI has been widely used to study older children, adolescents, and adults to detect brain functional development. However, young children, especially newborns, infants, and toddlers, are less likely to perform specific tasks. Therefore, resting-state fMRI (rs-fMRI) remains an indispensable tool, as it allows researchers to avoid having children perform specific tasks \citep{Zhang2019}. The BCP provides unprecedented high-resolution rs-fMRI and a dedicated processing pipeline, greatly assisting researchers in constructing functional networks and characterizing the longitudinal development of infant and toddler brain function.  

Using rs-fMRI, we can explore the developmental process of infant brain function which develops extremely rapidly and exhibits significant differences compared to adults, particularly in terms of the longitudinal dynamic segregation and integration of brain networks characterized by the evolution of modular structure \citep{Zhang2019}. The modular structure reflects how the brain regions (network nodes) are allocated to different brain subsystems. Besides, Changes of FC in the infant brain can reflect the rapid development of behavior and cognitive functions at different age stages \citep{Damoiseaux2009}. Analyzing cross-sectional data allows researchers to compare infant FC with another older cohort (e.g., adults) or perform longitudinal comparisons in later stages of development \citep{Gao2015}. 

The organization of brain function or cognition can be characterized by a multiscale hierarchical structure, extending from individual neurons and macrocolumns to larger, macroscopic brain regions \citep{Kringelbach2020, Park2013}. These FC patterns can vary significantly across different individuals \citep{Monti2017a, Betzel2019}. The variations in FC patterns arise not only from fluctuations in latent cognitive states, encompassing mental processes like thoughts, ideas, awareness, arousal, and vigilance \citep{Taghia2018a, Razi2017}, occurring unpredictably during both resting state \citep{Razi2015, Razi2016} and task-related brain activity \citep{Cribben2012, Gonzalez-Castillo2018, Vidaurre2018}. They are also influenced by non-neural physiological factors, including head motion, cardiovascular and respiratory effects, and noise stemming from hardware instability \citep*{Hutchison2013, Lurie2020}. 
The cognitive states and the presence of noise both change FC between pairs of brain areas, which may result in significant changes of the modular structure of FC \citep{Bassett2013, Cribben2017, Ting2021, Bian2021}.

A significant challenge lies in the unreliability of estimating modular structure using group averaged functional connectivity (FC). The first issue is that the group averaged FC ignores information that is shared across individuals \citep{Lehmann2021}. Another issue is that the outlier of some individual FCs may result in biased estimation for calculating group averaged FC. Therefore, it is very important to find a way to estimate the group-level modular structure to depict the subsystem of functional networks and quantify the inter-individual variability at each stage of infant brain development. 

In fMRI studies, several module-detection methods have been employed to characterize the subsystem of functional brain networks. For example, a stochastic block model with non-overlapping sliding windows was employed to characterize dynamic FC for networks. In this approach, edge weights were estimated by averaging coherence matrices across subjects, where a threshold was applied to binarize the matrices \citep*{Robinson2015}. However, this method might not preserve the full information of the time series and may overlook the assessment of inter-subject variability. A recent Bayesian inference-based method \citep{Bian2021} estimated the group averaged adjacency matrix, treating it as an observation that retains all information about the time series of the subjects. Nevertheless, this approach overlooked the variability in FC across different subjects and disregarded the higher-order topological properties of an individual's network. It utilised a Bayesian latent block model (LBM) with a Gaussian distribution to identify the modular structure, inferred using Markov chain Monte Carlo (MCMC) sampling \citep{Metropolis1953, Hastings1970}. The proposed model included a pre-determined number of modules through model selection. One problem of the methods based on block model and MCMC is that the estimation performance for high network dimensionality is limited, especially for brain networks with large number of regions.

In order to reliably identify the modular structure of FC, recent studies have shifted their attention to multiple networks across various subjects. They employ group-level analysis to estimate common features of network patterns. One method based on a multi-subject stochastic block modelling can flexibly assess the inter-individual variations \citep{Pavlovic2020}, but it also treats the FCs as binary edges. Another method for multi-subject analysis is based on multilayer modularity \citep*{Bassett2011, Bassett2013}, where a modularity quality function is optimized to partition the functional networks into subsystems with different modular resolutions. Alternative approaches capable of capturing the dynamics of brain networks at both individual and group levels, while considering between-subject variations in BOLD time series include \cite*{Ting2021} and \cite{Betzel2019}. All of the above module detection methods were applied to the studies of functional networks of adults. For module detection in infant studies, \cite{wen2019} depicted the first-year development of modules in infant brain functional networks with 3 month interval and reported an increasing number of functional modules during the infant development. However, one limitation about this method is that only five highest or weakest subjects were adopted for calculating the final group-level FC, which may lose large amounts of information contained in the group. 

In this paper, we capture the modular structure of group representative network based on multilayer module detection method, which is able to retain the variability across subjects for infant functional module development. Specifically, for individual-level FC, we use the maximization of modularity quality function to estimate modular structure quantified by module labels with different modularity resolutions \citep{Bassett2011, Bassett2013}. In the maximization of modularity quanlity function, we want to find the best partition of the network, such that there are more edges within the module than the expected number of edges by chance. For group representative network, we model the estimated individual-level modules by a Bayesian method based on Categorical-Dirichlet conjugate pair and define a maximum label assignment probability matrix (MLAPM) providing information about the group-level modules. For module development, we define a Jaccard similarity coefficient for evaluating the evolution of modules for the longitudinal development of infant functional networks. Compared with the aforementioned conventional group averaging methods, the method based on Bayesian modelling is more robust with respect to the stationarity of the module evolution for different modularity resolutions. Finally, we analyse the infant functional module development using our proposed method and compare the module similarity between sleeping and awake conditions for both female and male infants. We discovered that the female infant demonstrated more distinctive modules between sleeping and awake conditions, which may imply their relative quiet and restful sleep compared with male infant.

\section{Methods}

\begin{figure*}[!ht]
\centering
\includegraphics[width=1\linewidth]{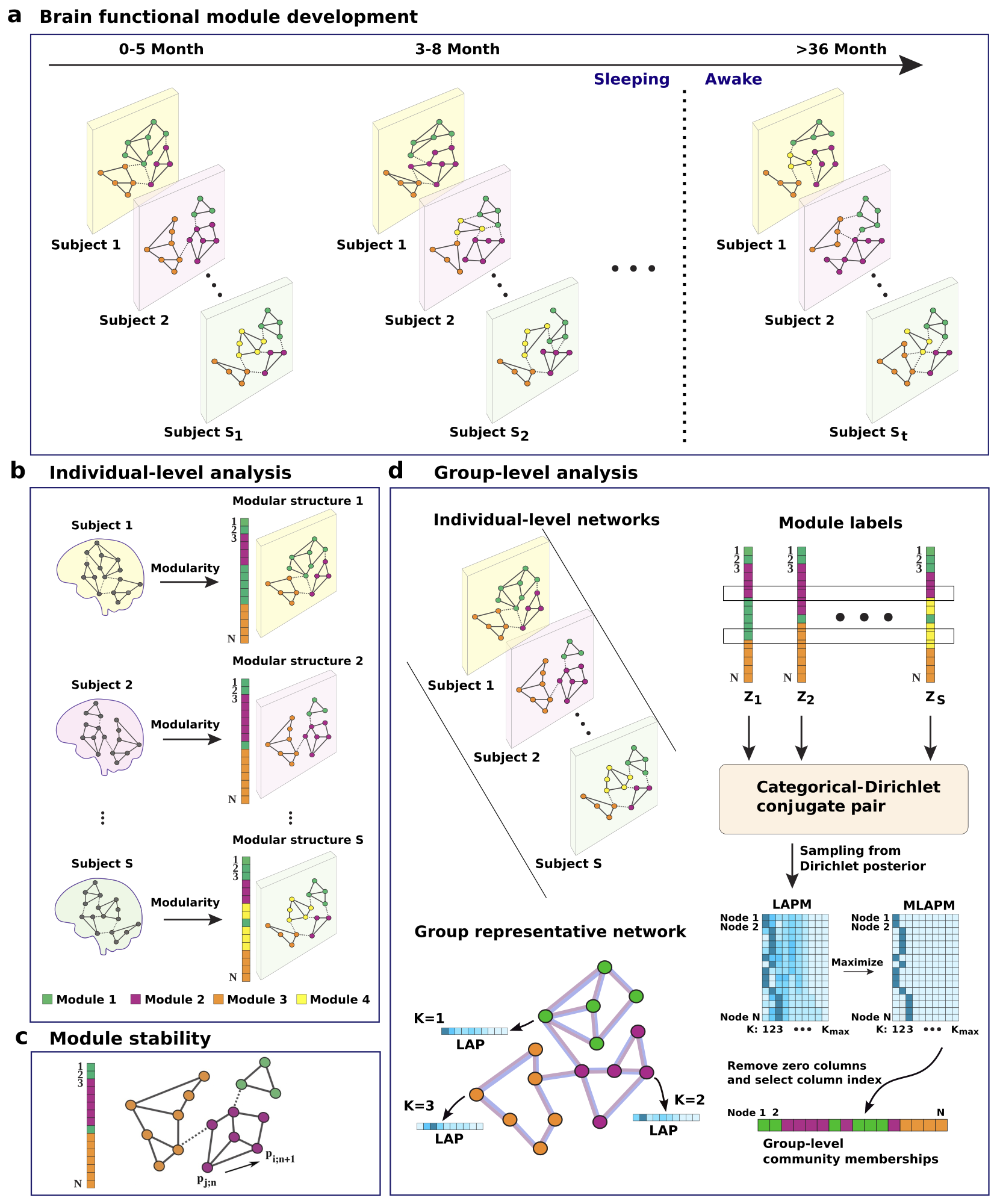}
\caption{\footnotesize \textbf{Overview of the framework.} \textbf{a} The brain functional module development. The subjects were divided into different age ranges, where the subjects under 36 month were scanned under sleeping condition and those above 36 months were scanned under awake condition. Different colours of nodes represent different module labels. \textbf{b} Individual-level analysis. The FC is computed via Pearson's correlation from the BOLD signals extracted from brain regions of interest (ROIs). The module labels of the network are estimated by maximizing modularity quality function for each subject. Nodes belonging to the same module are shown with the same colour. \textbf{c} The module stability. A random walker jumps around the network. For example, the random walker is located at node j with probability $p_{j;n}$ at time step n, then jumps to node i with probability $p_{i;n+1}$ at time step n+1. \textbf{d} Group-level analysis. We examine the group representative network, incorporating data from all subjects (from subject 1 to $S$). The module labels $\lbrace\bm{z}_{1},\cdots,\bm{z}_{s}\rbrace$ estimated for each individual are regarded as realizations of an underlying group-level modular structure. 
We employ a Bayesian model with a categorical-Dirichlet pair to model the module labels. Subsequently, we calculate a Maximum Label Assignment Probability Matrix (MLAPM) by maximizing each row of a Label Assignment Probability Matrix (LAPM). The MLAPM encodes the information about the modules of the group representative network, with each row representing a vector of Label Assignment Probabilities (LAP) for a specific node.}
\label{schematic_modelling}
\end{figure*}

The segregation and integration of functional networks change rapidly during the infant brain development (Fig.\ref{schematic_modelling}a). Suppose that the participants are divided into different age brackets (e.g. 0-5 month, 3-8 month etc), and there are different numbers of subjects within each age range (e.g. $S_{1}$ in 0-5 month, $S_{2}$ in 3-8 month, $...$, and $S_{t}$ for infants older than 36 month). In this paper, we develop a novel multilayer modular structure detection method based on modularity at the individual level (Fig.\ref{schematic_modelling}b) and Bayesian modelling at the group level (Fig.\ref{schematic_modelling}d) to robustly estimate the modular structure of the group representative network, meanwhile accounting for inter-individual variability within each age range. We will first illustrate how to model the FC of each subject with modularity \citep{Newman2004, Newman2006} and estimate the individual-level module labels by maximizing the modularity quality function (Fig.\ref{schematic_modelling}b). Then, we will elaborate the utilization of conjugate Bayesian pairs for modelling the estimated module labels at the individual level. We conduct parameter inference by sampling from posterior density to characterize the group-level modular structure (Fig.\ref{schematic_modelling}d). Note that in Bayesian inference, the posterior density results from the combination of prior and likelihood. If both the prior and posterior are derived from the same distribution family, we refer to the prior as a conjugate for the likelihood, forming a prior-likelihood conjugate pair.

\subsection{Individual-level modelling of the modular structure}

In this work, the modular structure characterizes how the brain regions are allocated to different subsystems. Each module represents a specific brain subsystem. For each individual functional network, the modular structure is detected by the maximization of its modularity which partitions a network into modules such that there are more densely connected edges within the modules than the edges expected by chance. 

We first illustrate the modularity quality function $Q$ \citep{Newman2006a} in terms of the module stability \citep{Delvenne2010, Lambiotte2008}. We elaborate the module stability  (Fig.\ref{schematic_modelling}c) as follows. Suppose that there is a random walker moving around the network and the discrete-time dynamics of the random walker follows 
\begin{equation}
p_{i;n+1}=\sum_{j}\frac{A_{ij}}{k_{j}}p_{j;n},
\end{equation}
where $p_{j;n}$ is the probability density of the random walker on node $j$ at step $n$, and $p_{i;n+1}$ is the probability density of the random walker moving from node $j$ to node $i$ at step $n+1$. $A_{ij}$ is the weight of connectivity going from $j$ to $i$, and $k_{j}=\sum_{i}A_{ij}$ is the strength of the connections going into node $j$. The stationary probability density of the random walker characterizes the chance of the random walker located at node $i$ at stationarity and it can be written as $p_{i}^{*}=k_{i}/2m$, where $m=\frac{1}{2}\sum_{i,j}A_{ij}$ is the sum of the edge strength of the network. The discrete module stability with time step one $R(1)$ characterizes the probability of a random walker to be in the same module during two successive time steps minus the probability for two independent walkers in the same module at stationarity, and it can be defined as
\begin{eqnarray}
R(1)&=&\sum_{i,j}[\frac{A_{ij}}{k_{j}}\frac{k_{j}}{2m}-\frac{k_{i}}{2m}\frac{k_{j}}{2m}]\delta(g_{i},g_{j})\\
&=&\frac{1}{2m}\sum_{i,j}[A_{ij}-\frac{k_{i}k_{j}}{2m}]\delta(g_{i},g_{j})\\
&=&Q,
\end{eqnarray}
the form of which is exactly the same as the modularity quality function Q. The delta function $\delta(g_{i},g_{j})=1$, if node $i$ and $j$ are in the same module ($g_{i}=g_{j}$), and 0 otherwise. 

\cite{Lambiotte2008} expanded the module stability to the continuous case, where the random walker jumps around the network with independent and identical homegeneous Poisson process governed by the normalized Laplacian dynamics \begin{eqnarray}
\dot p_{i}&=&\sum_{j}\frac{A_{ij}}{k_{j}}p_{j}+p_{i}.
\end{eqnarray}
Given the stationary probability density $p_{j}^{*}=k_{j}/2m$, the continuous module stability with time $t$ can be defined as
\begin{equation}
R(t)=\sum_{i,j}[(e^{t\bm{L}})_{ij}\frac{k_{j}}{2m}-\frac{k_{i}}{2m}\frac{k_{j}}{2m}]\delta(g_{i},g_{j}),
\end{equation}
where $L_{ij}=\frac{A{ij}}{k_{j}}-\delta_{ij}$.
If the linear approximation $(e^{t\bm{L}})_{ij}\approx \delta_{ij}+tL_{ij}$ is introduced, the quality function can be derived as
\begin{equation}
Q(t)=\frac{1}{2m}\sum_{i,j}[tA_{ij}-\frac{k_{i}k_{j}}{2m}]\delta(g_{i},g_{j}).
\end{equation}
Dividing by $t$ to this equation does not affect the optimization of the quality function, so the quality function can be rewritten with resolution parameter $\gamma=1/t$, such that
\begin{equation}
Q=\frac{1}{2m}\sum_{i,j}[A_{ij}-\gamma\frac{k_{i}k_{j}}{2m}]\delta(g_{i},g_{j}).
\end{equation}
In this work, the best partition of the network is estimated by maximizing the above modularity quality function $Q$, which is implemented by a MATLAB function \textit{modularity\_und.m} in Brain Connectivity Toolbox \citep{Rubinov2010}. The resolution of the modules which determines the mean module size and the number of modules is controlled by the resolution parameter $\gamma$. Larger value of $\gamma$ results in larger number of modules and the smaller mean module size.

For individual-level analysis in this work, we apply the modularity to each subject within each age range with different values of $\gamma$, resulting in coarse-to-fine partitions of network for each subject. In the next section, we will illustrate the group-level modelling and how to characterize the group-level modular structure in each age range. 

\subsection{Group-level analysis of the module development}

At the group level, we model the module labels of networks estimated from the individual-level analysis in each age range. We take the individual module labels as the observation for group-level analysis (see in Fig.\ref{schematic_modelling}d). Suppose that there are $S$ number of subjects (note that $S$ is different in different age ranges). We define a matrix 
\begin{equation}
\bm{Z}=\begin{pmatrix}
  z_{11} & \cdots & z_{1s} & \cdots & z_{1S}\\
  \vdots & \ddots & \vdots & \ddots & \vdots\\
  z_{i1} & \cdots & z_{is} & \cdots & z_{iS}\\
  \vdots & \ddots & \vdots & \ddots & \vdots\\
  z_{N1} & \cdots & z_{Ns} & \cdots & z_{NS}\\
\end{pmatrix}
\end{equation}
to represent the module labels for all of the subjects estimated using the modularity quality function, where $N$ is the number of nodes and $S$ is the number of subjects in a specific age range. Each row vector $\bm{z}_{i}$ contains the module labels of a specific node $i$ over all of the subjects within the age range, and each column vector $\bm{z}_{s}$ represents the module labels of a specific subject $s$ in the group. 

We model each row of $\bm{Z}$, namely $\bm{z}_{i}=(z_{i1},\cdots,z_{is},\cdots,z_{iS})$ for $S$ subjects for node $i$ using a categorical-Dirichlet conjugate pair. Each label $z_{is}$ follows a categorical distribution $z_{is}\sim \mbox {Categorical}(1;\bm{r}_{i})$, where $\bm{r}_{i}=(r_{i1},\cdots,r_{ik},\cdots,r_{iK})$ is a vector of label assignment probabilities (LAP) such that $\sum_{k=1}^{K} r_{ik}=1$, and $K$ is the maximum number of elements in $\bm{Z}$ in the group. 

We define a label assignment probability matrix (LAPM)
\begin{equation}
\bm{R}=\begin{pmatrix}
  r_{11} & \cdots & r_{1k} & \cdots & r_{1K}\\
  \vdots & \ddots & \vdots & \ddots & \vdots\\
  r_{i1} & \cdots & r_{ik} & \cdots & r_{iK}\\
  \vdots & \ddots & \vdots & \ddots & \vdots\\
  r_{N1} & \cdots & r_{Nk} & \cdots & r_{NK}\\
\end{pmatrix},
\end{equation}
and
\begin{equation}
p(z_{is}\vert\bm{r}_{i},K)=\prod_{k=1}^{K}r_{k}^{I_{k}(z_{is})}, \mbox{where\ }  I_{k}(z_{is})=
\begin{cases}
1, \ \mbox{if}\ z_{is}=k\\
0, \ \mbox{if}\ z_{is}\neq k\\
\end{cases}.
\end{equation}
Consider a prior of Dirichlet distribution $\bm{r}_{i}\sim \mbox{Dirichlet}(\bm{\alpha})$
\begin{equation}
p(\bm{r}_{i}\vert K)=N(\bm{\alpha})\prod_{k=1}^{K}r_{k}^{\alpha_{k}-1}, 
\end{equation}
with the normalization factor $N(\bm{\alpha})=\frac{\Gamma(\sum_{k=1}^{K}\alpha_{k})}{\prod_{k=1}^{K}\Gamma(\alpha_{k})}$. In this paper, we set $\alpha_{k}=1$.
The posterior can be expressed as
\begin{eqnarray}
p(\bm{r}_{i}\vert \bm{z}_{i},K)&\propto& \prod_{s=1}^{S} p({z}_{is}\vert \bm{r}_{i}) \times p(\bm{r}_{i})\\
&=&\prod_{s=1}^{S}\prod_{k=1}^{K}r_{k}^{I_{k}(z_{is})}\times N(\bm{\alpha})\prod_{k=1}^{K}r_{k}^{\alpha_{k}-1}\\
&=& N(\bm{\alpha})\prod_{k=1}^{K}r_{k}^{\sum_{s=1}^{S}I_{k}(z_{is})+\alpha_{k}-1}\\
&=&\frac{N(\bm{\alpha})}{N(\bm{\alpha}')}N(\bm{\alpha}')\prod_{k=1}^{K}r_{k}^{\sum_{s=1}^{S}I_{k}(z_{is})+\alpha_{k}-1},
\end{eqnarray}
where $\alpha_{k}'=\sum_{s=1}^{S}I_{k}(z_{is})+\alpha_{k}$, $N(\bm{\alpha}')=\frac{\Gamma(\sum_{k=1}^{K}(\sum_{s=1}^{S}I_{k}(z_{is})+\alpha_{k}))}{\prod_{k=1}^{K}\Gamma(\sum_{s=1}^{S}I_{k}(z_{is})+\alpha_{k})}$, and 
the posterior $\bm{r}_{i}\vert \bm{z}_{i}\sim \mbox{Dirichlet}(\bm{\alpha}')$.
The posterior for the network can be expressed as
\begin{equation}
p(\bm{R}\vert \bm{Z},K)=\prod_{i=1}^{N}p(\bm{r}_{i}\vert \bm{z}_{i},K).
\end{equation} 
We use the maximum LAPM (MLAPM), $\bm{R}_{\mbox{max}}$, the maximum probability at each row of $\bm{R}$, to provide the information of modular structure as shown in Fig.\ref{schematic_modelling}d. The column indices of $\bm{R}_{\mbox{max}}$ after removing all of the zero columns from the estimation of the group-level module labels denoted as $\bm{z}^{G}$. We do this modular structure estimation for the group representative network in each age range using the same process.
%%%%%%%%%%%%%%%%%%%%%%%%%%%%%%%%%%%%%%%%%%%%
\subsection{Evaluating module evolution}

In this section, we quantify the module evolution using Jaccard similarity coefficient $J(\bm{z}_{t}^{G},\bm{z}_{t+\tau}^{G})$ which is defined as 
\begin{equation}
J(\bm{z}_{t}^{G},\bm{z}_{t+\tau}^{G})=\frac{\vert\bm{z}_{t}^{G}\cap\bm{z}_{t+\tau}^{G}\vert}{\vert\bm{z}_{t}^{G}\cup\bm{z}_{t+\tau}^{G}\vert},
\end{equation}
where $\bm{z}_{t}^{G}$ is the group-level modular structure in the age range $t$ (e.g. $t=0$ for 0-5 month, $t=3$ for 3-8 month), and $\bm{z}_{t+\tau}^{G}$ is the modular structure for age range $\tau$ months later ($\tau$ takes the value of 6, 12, and 18).  Here, $\cap$ indicates the intersection of two vectors of module labels, and $\cup$ is the union. We also call Jaccard similarity coefficient as the $J$ value in the rest of paper. The larger $J$ value means that the modular structure is more similar between the two age ranges. For two consecutive non-ovelapping age ranges, smaller $J$ value means the module evolution of the infant brain development is relatively stationary.

\subsection{BCP fMRI data acquisition and preprocessing}

In this paper, we use the rs-fMRI data from the BCP \citep*{Howell2019} which involves the infants from 2 weeks to 72 months of age. Scans from 546 subjects are collected using a 3T Siemens Prisma MRI scanner with 32 channel head coils. In this work, the rs-fMRI with both the phase coding directions of anterior-to-posterior (AP) and posterior-to-anterior (PA) are utilized. The  protocol parameters for rs-fMRI include FOV: 208 mm$\times$208 mm, voxel size: 2mm isotropic, flip angle: 52$^{\circ}$, TE: 37ms, and TR: 800ms (see more detailed information about the parameters in \cite{Howell2019}). The rs-fMRI preprocessing \citep*{Zhang2019} includes head motion correction, distortion correction, anatomical registration, one-step resampling, and denoising \citep*{Heo2022}. The FC (with 100, 200, 300, and 400 ROIs) is constructed by calculating the correlation matrix of the extracted BOLD signals using the Schaefer's atlas \citep*{Schaefer2018}. The dataset was partitioned into different age ranges as shown in Table \ref{agewindows}. 

\begin{table}[!ht]
\caption{Age ranges with different numbers of scans.}
\centering
\begin{tabular}{|c|c|c|c|c|c|c|c|c|c|r|l|} \hline
 Age range (month) & 0-5 & 3-8 & 6-11 & 9-14 & 12-17 & 15-23 & 18-29 & 24-36 & >36  \\ \hline
Scans (AP) & 58 & 88 & 112 & 128 & 113 & 107 & 108 & 67 & 29  \\ \hline
Scans (PA) & 60 & 92 & 114 & 129 & 115 & 105 & 109 & 68 & 29  \\ \hline
\end{tabular}
\label{agewindows}
\end{table}

\section{Experiments and results}

 We first evaluate the maximization of modularity quality function with different resolutions $\gamma$ using three metrics including modularity quality $Q$, the number of communities $K$, and the mean community size $S$. Secondly, we evaluate the performance of Bayesian modelling at the group level with respect to different values of resolution parameter $\gamma$. We compare the results of using Bayesian modelling and conventional group averaging methods for group-level analysis in terms of the module evolution of functional networks. Finally, we compare the module evolution between female and male infants under sleeping and awake conditions.

\subsection{Individual-level analysis based on modularity}

\begin{figure*}[!ht]
\centering
\includegraphics[width=1\linewidth]{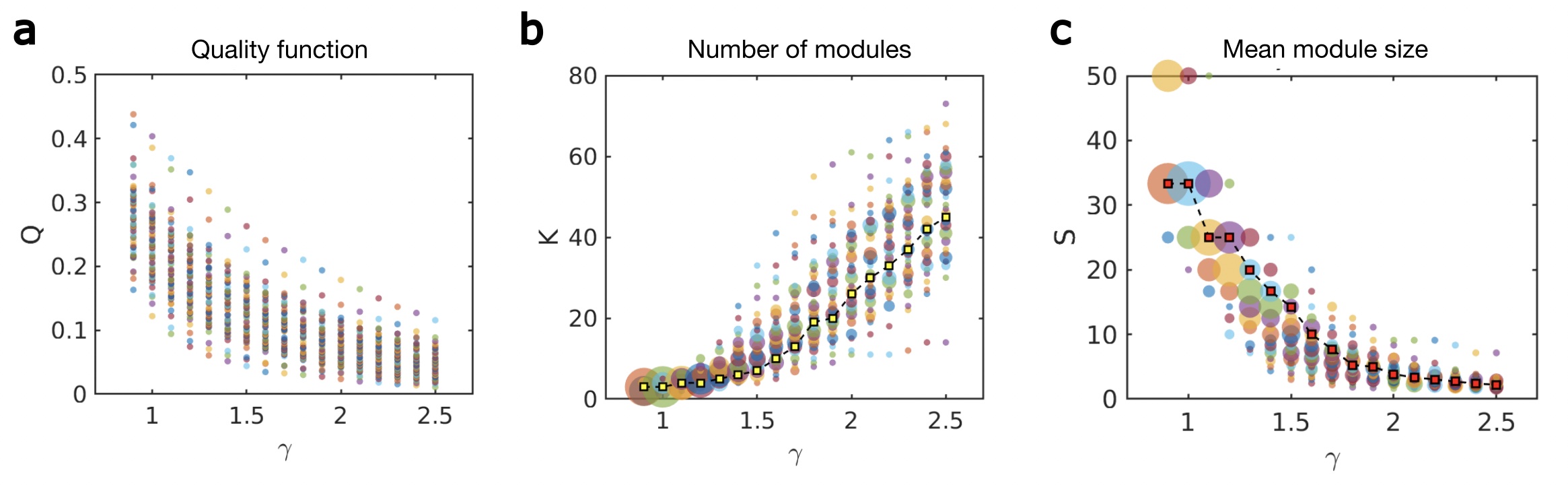}
\caption{\footnotesize \textbf{Modularity for individual analysis}. Modularity metrics including quality function $Q$, number of modules $K$, and mean module size $S$ for network of 100 ROIs (AP: the scan direction from anterior to posterior). \textbf{a} The modularity quality function $Q$ versus resolution parameter $\gamma$. Different colour dots represent different subjects. \textbf{b} The number of modules $K$ versus $\gamma$. The size of the dot indicates the number of subjects (larger dot size corresponds to larger number of subjects). The dashed curve with yellow square represents the estimated number of modules by Bayesian inference. \textbf{c} The mean module size $S$ versus $\gamma$. The dots have the same meaning as those in \textbf{b}. The dashed curve with red square represents the estimated mean module size by Bayesian inference.}
\label{modularity_metrics}
\end{figure*}
For individual-level analysis, we apply the maximization of modularity quality function to each subject in each age range. We visualize the quality function $Q$ for each subject with different values of resolution parameter $\gamma$ ($\gamma=0.9$, 1.0, 1.1, ...,  to 2.5) for a specific age range (AP, with 100 ROIs) as shown in Fig.\ref{modularity_metrics}a, where different colour dots represent different subjects. We can see that the $Q$ decreases with the increasing value of $\gamma$. We also plot the number of modules $K$ and the mean module size $S$ against $\gamma$. The $K$ decreases (Fig.\ref{modularity_metrics}b) and the $S$ increases (Fig.\ref{modularity_metrics}c) against the increasing value of $\gamma$.

\subsection{Group-level analysis based on Bayesian modelling}

\begin{figure*}[!ht]
\centering
\includegraphics[width=0.95\linewidth]{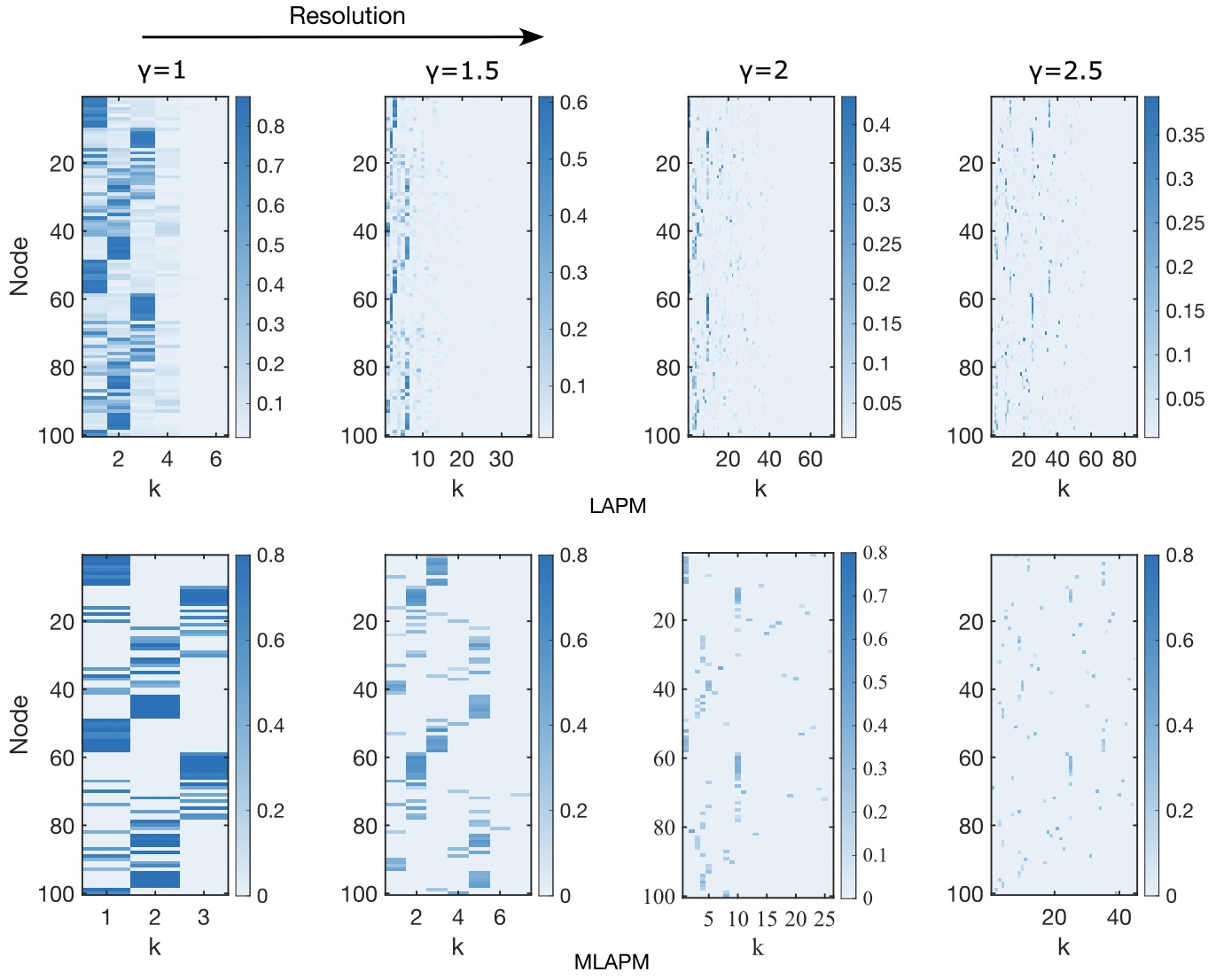}
\caption{\footnotesize \textbf{Estimation of module labels with different resolution parameters}. The module labels are first inferred separately for each subject using maximization of modularity quality function with different resolutions (e.g. from $\gamma= 0.9$, 1.0, 1.1, to 2.5, we just show the results of $\gamma= 1$, 1.5, 2, and 2.5 in the main text). Then we model the module labels of each node across all of the subjects at group level, where an LAPM is estimated from a Dirichlet posterior density for different values of $\gamma$ as shown in the upper panel. The elements in each row of the matrix represent the probabilities of a node assigned to different modules. Only the largest probability in each row of the LAPM is retained in the MLAPM and the columns with all zero values are removed (lower panel). The column index in MLAPM corresponds to the module labels at the group level. The color bars indicate the label assignment probability in the LAPM and the MLAPM.}
\label{resolution_analysis}
\end{figure*}

We visualize the estimation of group-level modular structure based on Bayeisan modelling with different values of $\gamma$ for a specific age range (e.g. age range 0-5 month) as shown in Fig.\ref{resolution_analysis}. At the group level, categorical-Dirichlet conjugate pairs are used to model the module labels of each row vector in $\bm{Z}$ estimated at the individual level. The module labels of each node across all of the subjects in a specific age range are regarded as the observations of categorical-Dirichlet conjugate pair (see Fig. \ref{schematic_modelling}). An LAPM is inferred by sampling from a Dirichlet posterior density (the upper panel of Fig.\ref{resolution_analysis}). Subsequently, an MLAPM is formed, encompassing information about the group-level modular structure. This is achieved by preserving the maximum probability within each row of LAPM, signifying the highest likelihood of the node being assigned to a particular module (the lower panel of Fig.\ref{resolution_analysis}). And finally, the zero columns are removed, resulting in the final version of MLAPM summarizing the group-level modular structure, where the number of columns represents the estimated number of modules $K$ at the group level. The row index denotes the node number, while the column index signifies the module labels at the group level. The bar indicates the value of the maximum assignment probability for the labels.

The module labels of different age ranges are inconsistent with each other due to the label-switching phenomenon. Here, we used the relabelling algorithm \citep{Nobile2007,Wyse2012} based on minimization of label vector distance and square assignment algorithm \citep{Carpaneto1980} to obtain a switched module labels cross different age ranges. 

\subsection{The module evolution of infant functional brain networks}

\begin{figure*}[!ht]
\centering
\includegraphics[width=1\linewidth]{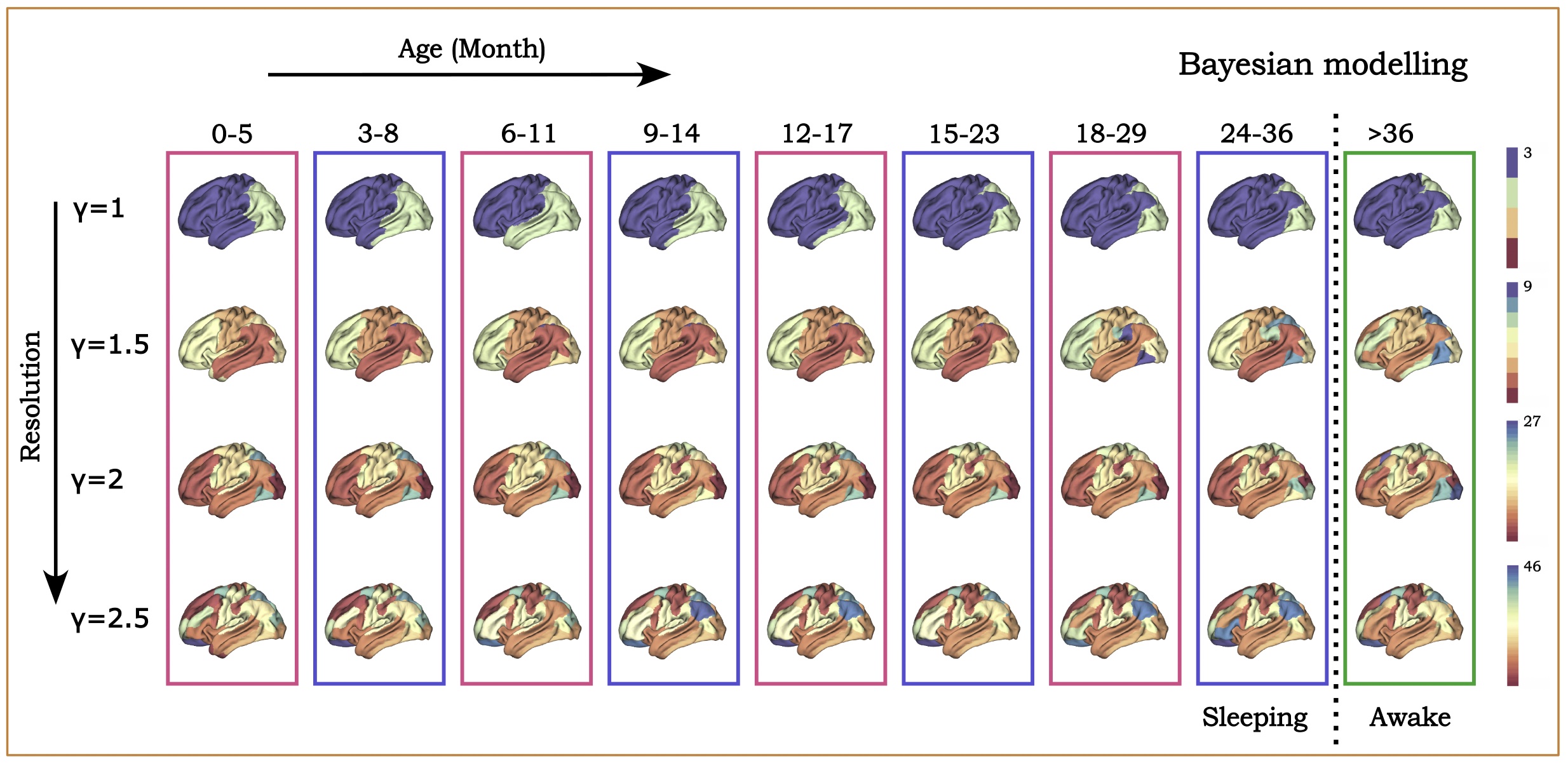}
\caption{\footnotesize \textbf{The development of modular structure of infant functional brain networks with different modularity resolutions}. The subjects are divided into different age ranges (from 0-5 month, 3-8 month, ..., and > 36 months). The infants before 36 month (pink and blue frames, and the same color indicates non-overlapping age ranges) are scanned under sleeping condition and those older than 36 month (green frame) are scanned under awake condition.}
\label{longitudinal_modular_development}
\end{figure*}

We illustrate the development of modular structure in infant functional brain networks at various modularity resolutions, as depicted in Fig. \ref{longitudinal_modular_development}. The modular structure in each age range is estimated based on Bayesian modelling and visualized using BrainSpace \citep{VosdeWael2020}. Here, all the age ranges with pink frame or blue frame are non-overlapping, where the subjects are under sleeping condition. The modular structure in the green frame corresponds to the subjects scanned older than 36 months under awake condition.

\begin{figure*}[!ht]
\centering
\includegraphics[width=0.9\linewidth]{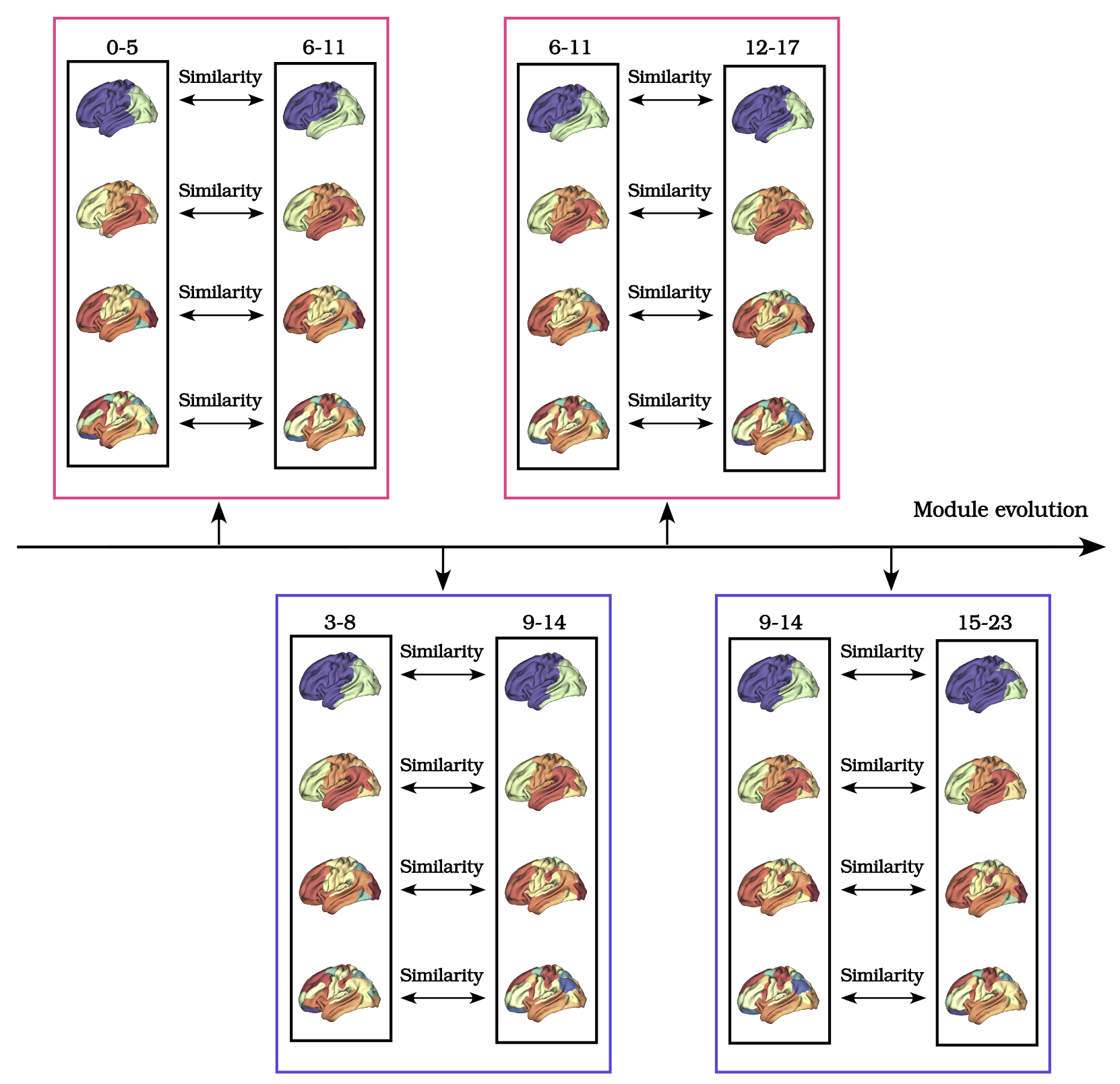}
\caption{\footnotesize \textbf{The module evolution with different modularity resolutions}. The module evolution is quantified by the Jaccard similarity coefficient of the modular structure between consecutive but non-overlapping age ranges (the similarity comparisons of the frames with same colour).}
\label{modular_evolution}
\end{figure*}

In this work, we are interested in how the modular structure evolves longitudinally. For that purpose, we evaluate the module evolution by assessing the Jaccard similarity between the neighbouring non-overlapping age ranges for each setting of modularity resolution $\gamma$ (Fig.\ref{modular_evolution}). 

\begin{figure*}[!ht]
\centering
\includegraphics[width=0.9\linewidth]{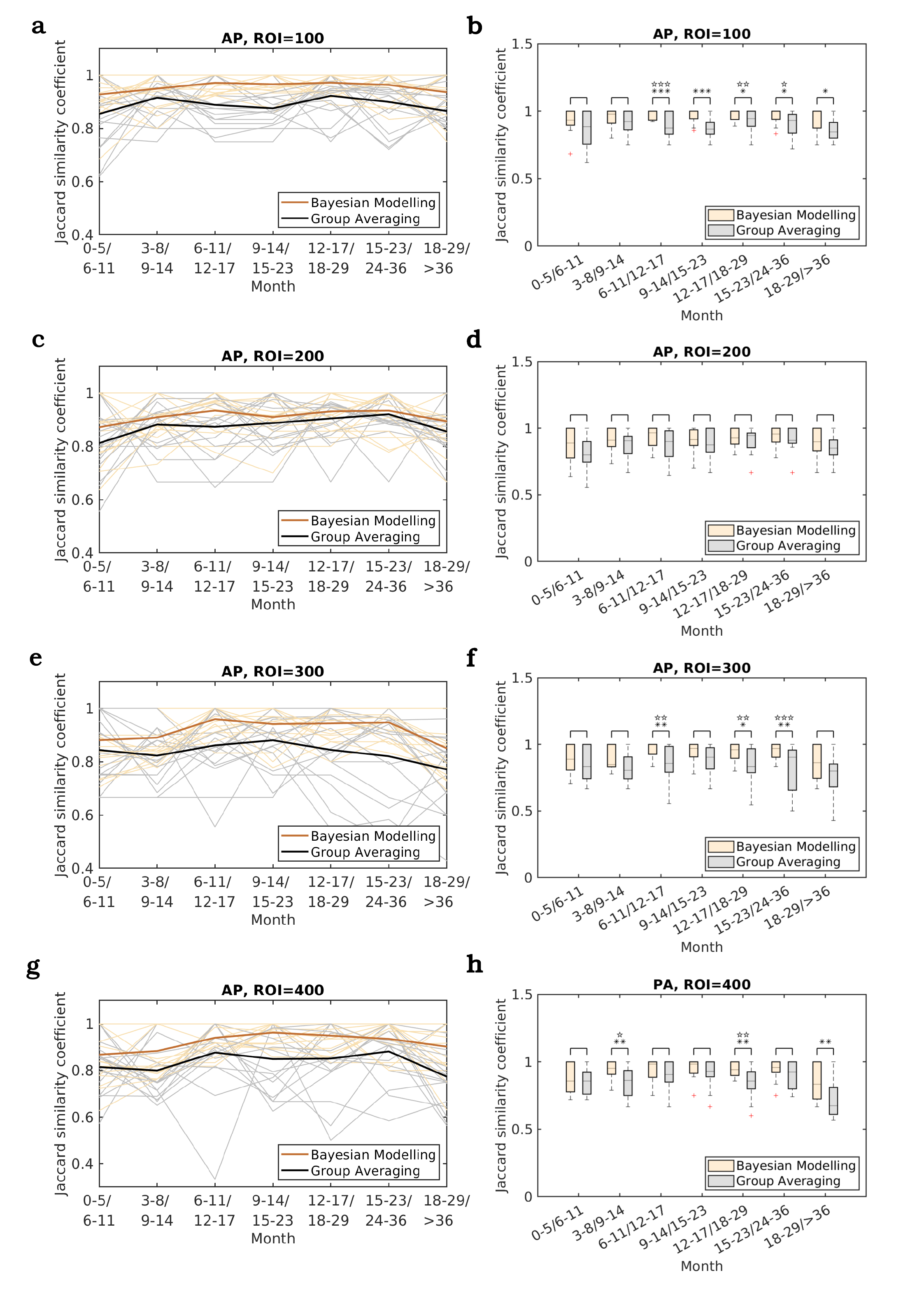}
\caption{\footnotesize \textbf{The comparison of the module evolution evaluated by our proposed Bayesian modelling method and conventional group averaging method (AP)}. The module evolution is evaluated for the infant brain networks with different numbers of ROIs (100 ROIs in \textbf{a} and \textbf{b}; 200 ROIs in \textbf{c} and \textbf{d}; 300 ROIs in \textbf{e} and \textbf{f}; and 400 ROIs in \textbf{g} and \textbf{h}). In the left panel, the light orange curves are the Jaccard similarity coefficients with different modularity resolutions $\gamma$ evaluated by Bayesian modelling method and grey lines are those evaluated by group averaging method. The darker orange curve and black curve are the mean values averaged across different values of $\gamma$. The boxplots in the right panel are the statistical analysis of the two methods. The star indicates the statistical significance of F-test (\ding{73}: $0.01<p<0.05$, \ding{73}\ding{73}: $0.001<p<0.01$, and \ding{73}\ding{73}\ding{73}: $p<0.001$) and the asterisk indicates the statistical significance of t-test ($\ast$: $0.01<p<0.05$, $\ast$ $\ast$: $0.001<p<0.01$, and $\ast$ $\ast$ $\ast$: $p<0.001$).}
\label{Jaccard_similarity_evolution}
\end{figure*}

We compared the performance of our proposed Bayesian modelling method for group level analysis with the conventional group averaging method in terms of the Jaccard similarity coefficient for different scales of functional networks with ROIs of 100, 200, 300, and 400 respectively as shown in Fig.\ref{Jaccard_similarity_evolution} (note that we demonstrate the results of AP in the main text, see the results of PA in the SI Fig.1). According to the results of the Jaccard similarity with different values of $\gamma$, the module evolution evaluated by Bayesian modelling demonstrates overall larger mean $J$ value compared with group averaging method (Fig.\ref{Jaccard_similarity_evolution}a). We see the similar results with respect to the mean $J$ values in Fig.\ref{Jaccard_similarity_evolution}c, e, and g as well. According to the results of t-test, the $J$ values of using Bayesian modelling are significantly larger than group averaging method in age ranges between 6-11 month and 12-17 month, 9-14 month and 15-23 month, 12-17 month and 18-29 month, 15-23 month and 24-36 month, and 18-29 month and >36 month. More importantly, the $J$ values of using Bayesian modelling show significantly less diversity compared with those of using group averaging method according to the F-test between the age ranges of 6-11 month and 12-17 month, 12-17 month and 18-29 month, and 15-23 month and 24-36 month (Fig.\ref{Jaccard_similarity_evolution}b), which means that the J values evaluated by Bayesian modelling are relatively stationary compared with those of group averaging method. We can see similar statistically significant results of the comparison between some age ranges using networks with ROIs of 200, 300 and 400 as well. The Bayesian modelling method demonstrates relatively stationary module evolution, implying a more robust group-level modular detection compared with that of the group averaging method.

\begin{figure*}[!ht]
\centering
\includegraphics[width=0.9\linewidth]{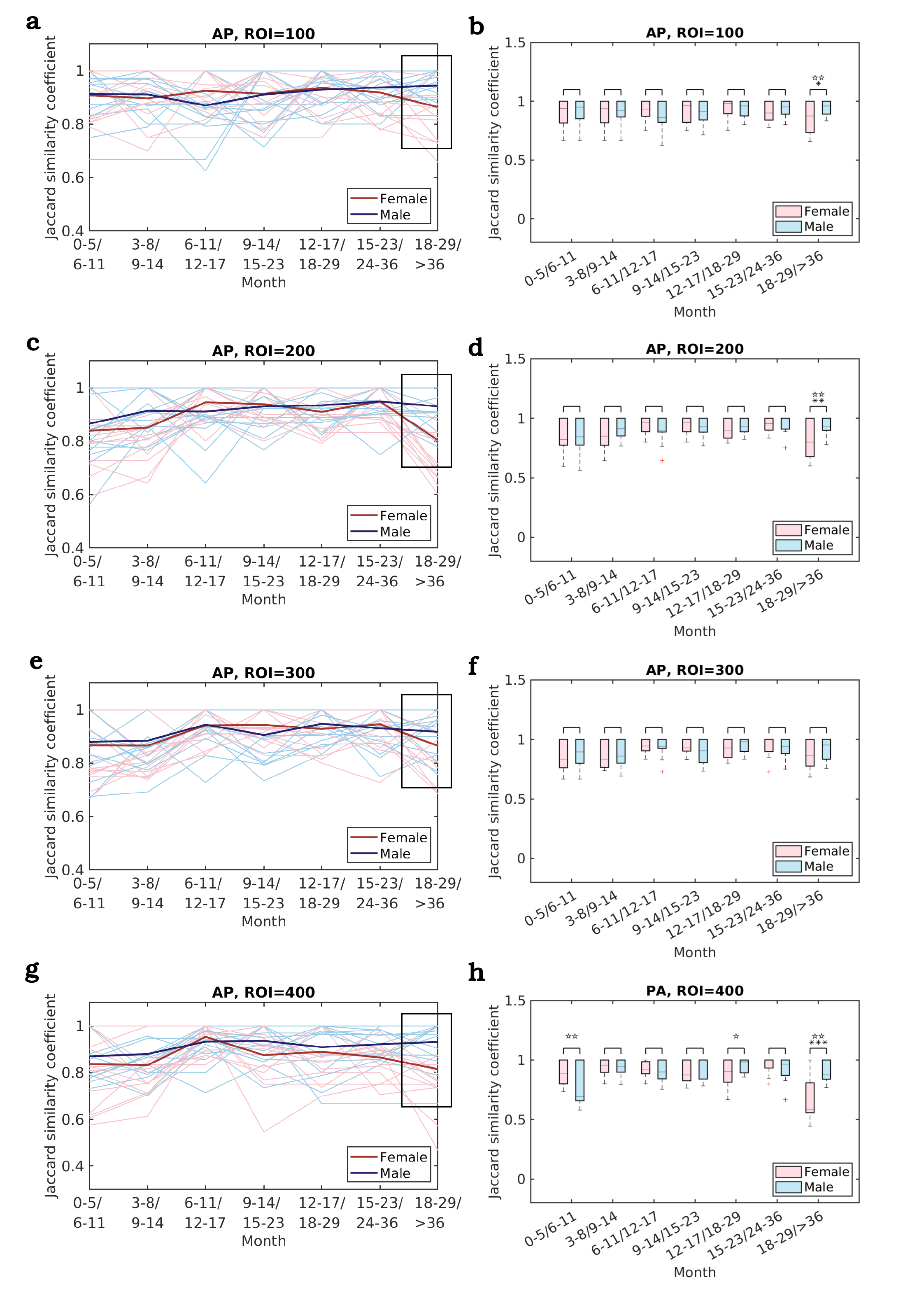}
\caption{\footnotesize \textbf{The comparison of the module evolution of female and male (AP)}. The module evolution is evaluated for the infant brain networks with different numbers of ROIs (100 ROIs in \textbf{a} and \textbf{b}; 200 ROIs in \textbf{c} and \textbf{d}; 300 ROIs in \textbf{e} and \textbf{f}; and 400 ROIs in \textbf{g} and \textbf{h}). In the left panel, the pink curves are the Jaccard similarity coefficients with different modularity resolutions $\gamma$ of female and light blue curves are those of the male. The darker pink curve and darker blue curve are the mean values averaged across different values of $\gamma$. The boxplots in the right panel are the statistical analysis of female and male. The star indicates the statistical significance of F-test (\ding{73}: $0.01<p<0.05$, \ding{73}\ding{73}: $0.001<p<0.01$, and \ding{73}\ding{73}\ding{73}: $p<0.001$) and the asterisk indicates the statistical significance of t-test ($\ast$: $0.01<p<0.05$, $\ast$ $\ast$: $0.001<p<0.01$, and $\ast$ $\ast$ $\ast$: $p<0.001$).}
\label{Jaccard_similarity_gender}
\end{figure*}

\subsection{Comparing the module evolution of female and male}

Next, we compare the module evolution between female and male, the results of which with the data of AP are shown in Fig.\ref{Jaccard_similarity_gender} (see the results of PA in SI Fig.2). There are no significant differences between female and male with respect to the module evolution under sleeping conditions. However, we surprisingly discovered that the female infant demonstrates significantly lower $J$ value between 18-20 month (sleeping condition) and >36 (awake condition) compared with that of the male infant (see the black rectangles in the plots of the left panel). 

\begin{figure*}[!ht]
\centering
\includegraphics[width=0.9\linewidth]{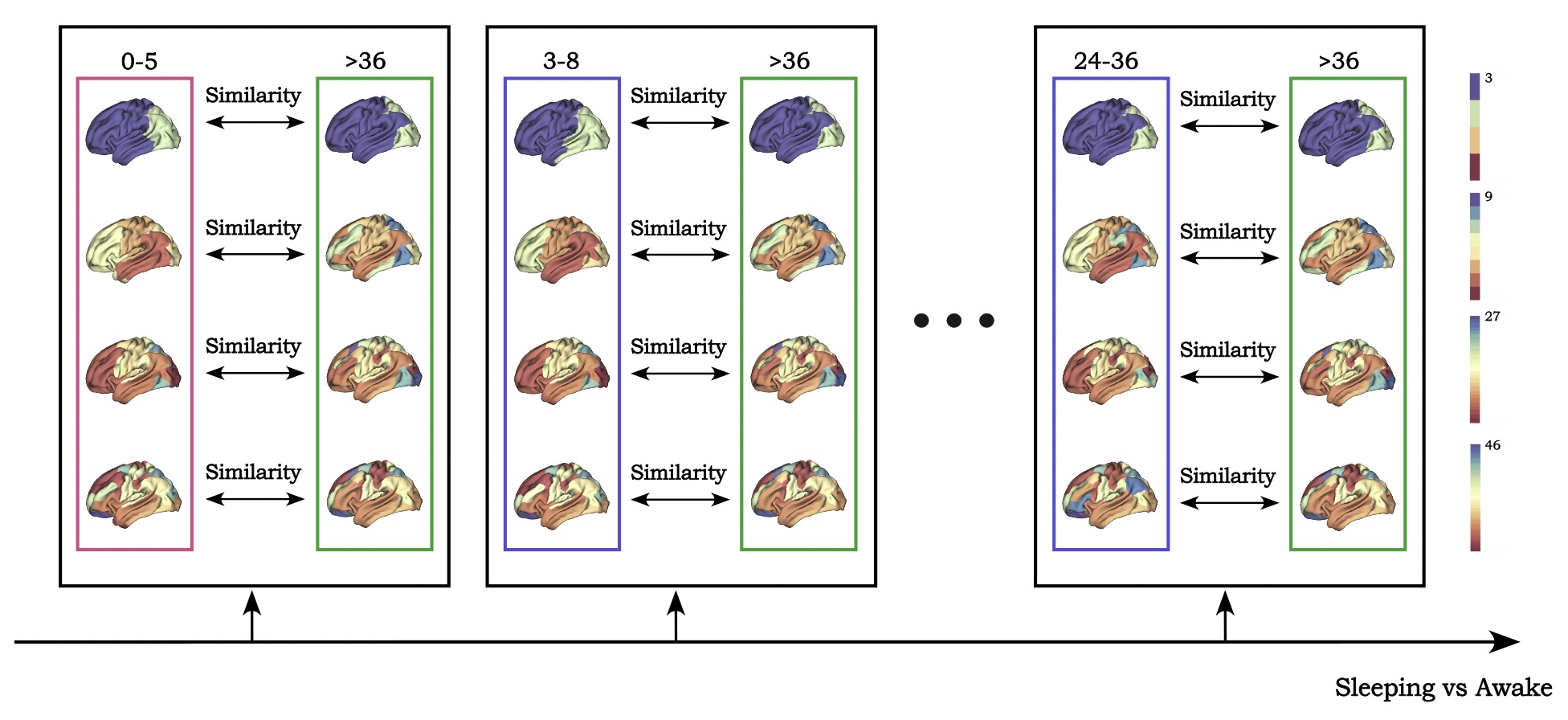}
\caption{\footnotesize \textbf{Modular structure comparison between sleeping and awake conditions}. We compare the modular structure between each age range (those before 36 month) and that of the age older than 36 month in terms of the Jaccard similarity.}
\label{modular_awake_sleep}
\end{figure*}

In order to gain an insight into the difference of modular structures between female and male under sleeping and awake conditions. We evaluated the Jaccard similarity between each age range under sleeping condition and under awake condition (Fig.\ref{modular_awake_sleep}). We then plot the corresponding $J$ values for female and male as shown in Fig.\ref{Jaccard_similarity_awake_sleep} (see the results of PA in SI Fig.3). We found that the female demonstrates smaller Jaccard similarity between sleeping (several age ranges) and awake conditions compared with that of the male. While, the male shows similar modular structure between sleeping and awake conditions. Regarding this results, one assumption is that male infants sleep less restful compared with female infants. Male infants may undergo more somatic movement or their brain may be more activated during sleeping condition, resulting in more similar modular structure between sleeping and awake conditions. We will discuss this aspect further in the next section.

\begin{figure*}[!ht]
\centering
\includegraphics[width=0.9\linewidth]{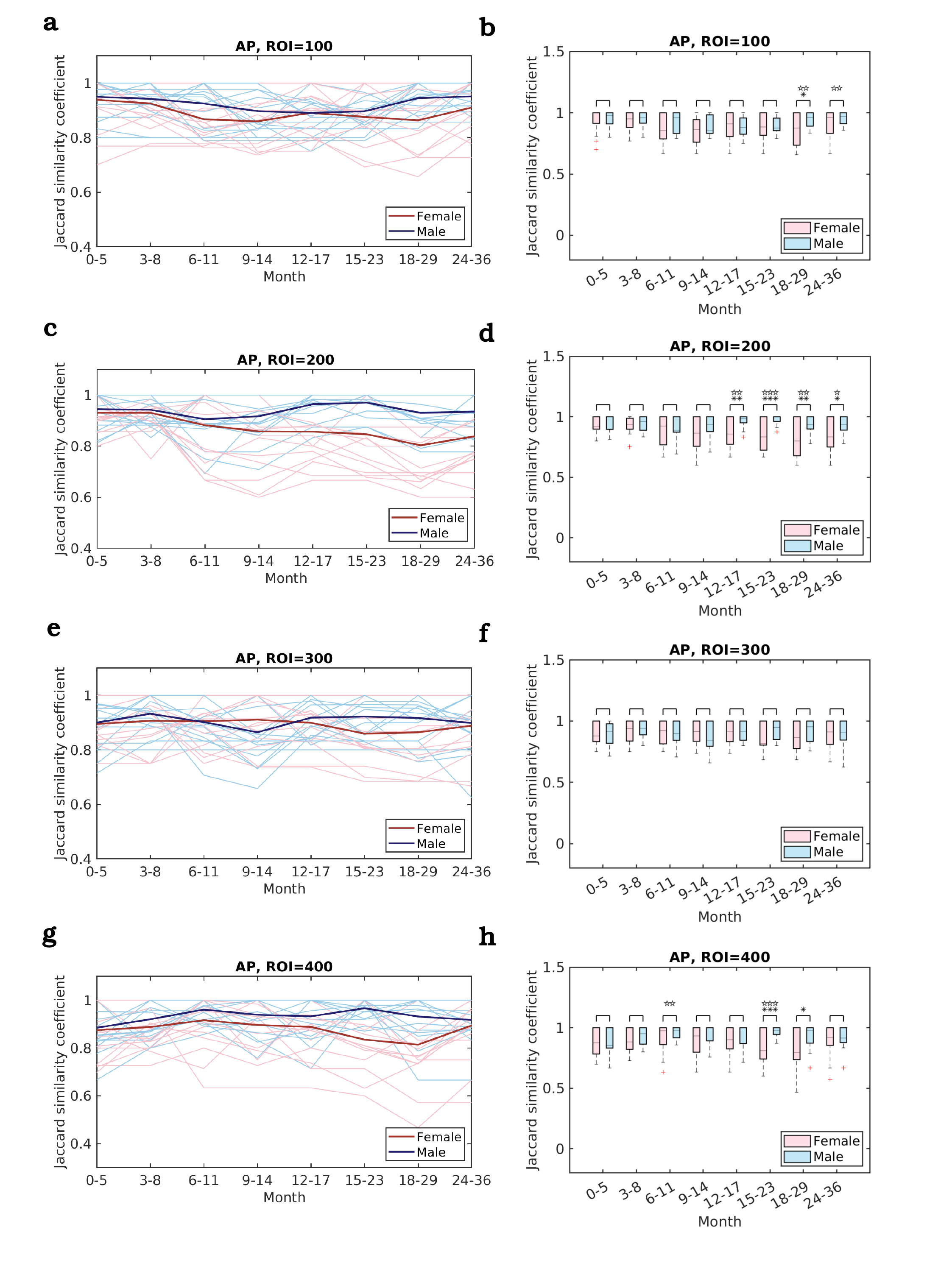}
\caption{\footnotesize \textbf{The similarity of modular structure between sleeping and awake conditions for female and male (AP)}. The Jaccard similarity between sleeping and awake conditions are evaluated for the infant brain networks with different numbers of ROIs (100 ROIs in \textbf{a} and \textbf{b}; 200 ROIs in \textbf{c} and \textbf{d}; 300 ROIs in \textbf{e} and \textbf{f}; and 400 ROIs in \textbf{g} and \textbf{h}). In the left panel, the pink curves are the Jaccard similarity coefficients with different modularity resolutions $\gamma$ of female and light blue curves are those of the male. The darker pink curve and darker blue curve are the mean values averaged across different values of $\gamma$. The boxplots in the right panel are the statistical analysis of female and male. The star indicates the statistical significance of F-test (\ding{73}: $0.01<p<0.05$, \ding{73}\ding{73}: $0.001<p<0.01$, and \ding{73}\ding{73}\ding{73}: $p<0.001$) and the asterisk indicates the statistical significance of t-test ($\ast$: $0.01<p<0.05$, $\ast$ $\ast$: $0.001<p<0.01$, and $\ast$ $\ast$ $\ast$: $p<0.001$). Note that the x-axis represents the age ranges of only sleeping condition, which is different from those in Fig.\ref{Jaccard_similarity_gender}.}
\label{Jaccard_similarity_awake_sleep}
\end{figure*}

\section{Discussion}

Many studies construct group-level functional networks by estimating group averaged FC \citep{Robinson2015, Bian2021} which ignores the difference and variation of networks across individuals. Other approaches, such as those that establish a threshold for FC \citep{Achard2006}, may lead to the loss of crucial topological information within the networks. Only a single observation (the group averaged FC) being taken into account may result in biased estimation of modular structure. Modelling the individual FC rather than the group averaged FC provides insight into the variability of both the observations themselves, and the variability in the modular structure between subjects. In this paper, we developed a novel multilayer module detection method based on Bayesian modelling for group representative networks in infant functional brain development. The segregation and integration of functional networks are defined by the underlying modular structure \citep{Bassett2011,Bassett2013,Betzel2018}. We evaluated the module evolution of infancy using Jaccard similarity and compared our proposed method to the conventional group averaging method.

According to the results, our method is more robust compared to the group averaging method in terms of the stationarity of module evolution. We first applied the maximization of modularity quality function for individual functional networks in each age range with different modularity resolutions. To estimate the modular structure of the group representative network, we introduced a Bayesian modelling approach based on conjugate analysis for determining module labels at the group level. This framework not only identifies the shared modular structure across individuals but also assesses the variability in modular structure patterns between individuals. 

In this paper, we relax the assumption of fixing the number of modules $K$ and assume that it is a variable which is controlled by the modularity resolution parameter. 
In this scenario, the inference of module labels are not restricted by a pre-determined number of modules. This enhances the flexibility of individual-level label estimation compared to methods that employ a fixed value of $K$. Besides, we retain the variation of $K$ for each subject in each age range by using different modularity resolutions, such that coarse-to-fine modular structures are considered for evaluating the module evolution of infant functional brain development. 

We compared the module evolution between female and male infants under sleeping and awake conditions. Our results indicate that female infant demonstrates more distinctive modular structures between these two conditions compared with male infant. One previous study reported that in contrast to preterm girls, preterm boys exhibited significantly reduced sleep duration, a tendency for increased active sleep and decreased quiet sleep, more wakefulness after sleep onset, and a tendency for shorter longest sleep periods \citep{Foreman2008}. Another recent study indicated that the disparity in sleep duration between genders may manifest early in life, female infants and preadolescents tend to have longer sleep durations than their male counterparts \citep{Franco2020}. Female infants seemed to experience more restful sleep than male infants aligns well with earlier reports of heightened sleep disruption in male compared to female infants. Specifically, maternal perceptions of their infant's sleep patterns have noted more problematic crying and increased night awakenings in male infants \citep{Richardson2010}. The above studies, to some extent, support our discovery about the distinctive difference of modular structures between female and male under sleeping and awake conditions.

The method proposed in this paper solves many of the problems compared to the conventional method. Nevertheless, the present work still has certain limitations. While the multilayer modular detection performs effectively, identifying modules in real functional brain networks remains a challenge. The absence of a standard algorithm for general module detection stems from the assumption that network architectures in the real world are generated from diverse latent processes. Module detection through modularity relies on the information from the adjacency matrix, capturing FC between node pairs. However, it does not leverage metadata or features associated with the nodes. An additional study introduced by \cite{Hoffmann2020} identifies modules in networks solely by utilizing raw time series data from nodes without considering edge observations. While much of the previous research treat node attributes or metadata as the ground truth of modules, a recent study indicates that metadata differ from ground truth. Treating them as such can lead to significant theoretical and practical issues \citep{Peel2017}. In the future, it is worthwhile to explore the relationship between module detection and node metadata concerning the network structure. An intriguing direction for future research involves hierarchical modelling of group-level node metadata in conjunction with multilayer FC.

In this paper, we have not utilized the assessment of neurological and behavioral functions to evaluate the behavioral significance of individual differences in estimating the group-level modular structure of infant functional network development. In the future research, we plan to relate the BCP behavioural data \citep{Howell2019} including Mullen scales of early learning (MSEL) for evaluations of language, motor, and perceptual abilities, Minnesota Executive Function Scale (MEFS) for assessment of cognitive flexibility, and the Dimensional Joint Attention Assessment (DJAA) for characterizing dimensional ratings of Responding to Joint Attention (RJA) and reflect individual differences, to the module evolution. We will be able to explain how variation in individual behaviour measurements relate to variations in the modular structure of infant functional brain network development.

\section*{Code availability}
The code of this work is available at:
\href{https://github.com/LingbinBian/InfantModuleEvolution}{https://github.com/LingbinBian/InfantModuleEvolution}. 

\section*{Acknowledgements}

This work was supported in part by National Natural Science Foundation of China (grant numbers 62131015, 62250710165, U23A20295, and 32371154), the STI 2030-Major Projects (No. 2022ZD0209000), Shanghai Municipal Central Guided Local Science and Technology Development Fund (grant number YDZX20233100001001), Science and Technology Commission of Shanghai Municipality (STCSM) (grant number 21010502600), and The Key R$\&$D Program of Guangdong Province, China (grant numbers 2023B0303040001, 2021B0101420006).

\newpage

%\bibliography{modular_development}

\begin{thebibliography}{}

\bibitem[Achard et~al., 2006]{Achard2006}
Achard, S., Salvador, R., Whitcher, B., Suckling, J., and Bullmore, E. (2006).
\newblock {A resilient, low-frequency, small-world human brain functional
  network with highly connected association cortical hubs}.
\newblock {\em Journal of Neuroscience}, 26(1):63--72.

\bibitem[Bassett et~al., 2013]{Bassett2013}
Bassett, D.~S., Porter, M.~A., Wymbs, N.~F., Grafton, S.~T., Carlson, J.~M.,
  and Mucha, P.~J. (2013).
\newblock {Robust detection of dynamic community structure in networks}.
\newblock {\em CHAOS}, 23:13142.

\bibitem[Bassett et~al., 2011]{Bassett2011}
Bassett, D.~S., Wymbs, N.~F., Porter, M.~A., Mucha, P.~J., Carlson, J.~M., and
  Grafton, S.~T. (2011).
\newblock {Dynamic reconfiguration of human brain networks during learning}.
\newblock {\em PNAS}, 108(18):7641--7646.

\bibitem[Betzel et~al., 2019]{Betzel2019}
Betzel, R.~F., Bertolero, M.~A., Gordon, E.~M., Gratton, C., Dosenbach, N.~U.,
  and Bassett, D.~S. (2019).
\newblock {The community structure of functional brain networks exhibits
  scale-specific patterns of inter- and intra-subject variability}.
\newblock {\em NeuroImage}, 202(September 2018):115990.

\bibitem[Betzel et~al., 2018]{Betzel2018}
Betzel, R.~F., Medaglia, J.~D., and Bassett, D.~S. (2018).
\newblock {Diversity of meso-scale architecture in human and non-human
  connectomes}.
\newblock {\em Nature Communications}, 9(1).

\bibitem[Bian et~al., 2021]{Bian2021}
Bian, L., Cui, T., {Thomas Yeo}, B., Fornito, A., Razi, A., and Keith, J.
  (2021).
\newblock {Identification of community structure-based brain states and
  transitions using functional MRI}.
\newblock {\em NeuroImage}, 244(September):118635.

\bibitem[Carpaneto and Toth, 1980]{Carpaneto1980}
Carpaneto, G. and Toth, P. (1980).
\newblock {Algorithm 548: Solution of the assignment problem [H]}.
\newblock {\em ACM Transactions on Mathematical Software (TOMS)},
  6(1):104--111.

\bibitem[Cribben et~al., 2012]{Cribben2012}
Cribben, I., Haraldsdottir, R., Atlas, L.~Y., Wager, T.~D., and Lindquist,
  M.~A. (2012).
\newblock {Dynamic connectivity regression: determining state-related changes
  in brain connectivity}.
\newblock {\em NeuroImage}, 61:720--907.

\bibitem[Cribben and Yu, 2017]{Cribben2017}
Cribben, I. and Yu, Y. (2017).
\newblock {Estimating whole-brain dynamics by using spectral clustering}.
\newblock {\em Journal of the Royal Statistical Society. Series C (Applied
  Statistics)}, 66:607--627.

\bibitem[Damoiseaux and Greicius, 2009]{Damoiseaux2009}
Damoiseaux, J.~S. and Greicius, M.~D. (2009).
\newblock {Greater than the sum of its parts: A review of studies combining
  structural connectivity and resting-state functional connectivity}.
\newblock {\em Brain Structure and Function}, 213(6):525--533.

\bibitem[Delvenne et~al., 2010]{Delvenne2010}
Delvenne, J.~C., Yaliraki, S.~N., and Barahon, M. (2010).
\newblock {Stability of graph communities across time scales}.
\newblock {\em Proceedings of the National Academy of Sciences of the United
  States of America}, 107(29):12755--12760.

\bibitem[Eyre et~al., 2021]{Eyre2021}
Eyre, M., Fitzgibbon, S.~P., Ciarrusta, J., Cordero-Grande, L., Price, A.~N.,
  Poppe, T., Schuh, A., Hughes, E., O'Keeffe, C., Brandon, J., Cromb, D.,
  Vecchiato, K., Andersson, J., Duff, E.~P., Counsell, S.~J., Smith, S.~M.,
  Rueckert, D., Hajnal, J.~V., Arichi, T., O'Muircheartaigh, J., Batalle, D.,
  and Edwards, A.~D. (2021).
\newblock {The Developing Human Connectome Project: Typical and disrupted
  perinatal functional connectivity}.
\newblock {\em Brain}, 144(7):2199--2213.

\bibitem[Foreman et~al., 2008]{Foreman2008}
Foreman, S.~W., Thomas, K.~A., and Blackburn, S.~T. (2008).
\newblock {Individual and gender differences matter in preterm infant state
  development}.
\newblock {\em JOGNN - Journal of Obstetric, Gynecologic, and Neonatal
  Nursing}, 37(6):657--665.

\bibitem[Fox and Raichle, 2007]{Fox2007}
Fox, M.~D. and Raichle, M.~E. (2007).
\newblock {Spontaneous fluctuations in brain activity observed with functional
  magnetic resonance imaging}.
\newblock {\em Nature Reviews Neuroscience}, 8(9):700--711.

\bibitem[Franco et~al., 2020]{Franco2020}
Franco, P., Putois, B., Guyon, A., Raoux, A., Papadopoulou, M.,
  Guignard-Perret, A., Bat-Pitault, F., Hartley, S., and Plancoulaine, S.
  (2020).
\newblock {Sleep during development: Sex and gender differences}.
\newblock {\em Sleep Medicine Reviews}, 51:101276.

\bibitem[Gao et~al., 2015]{Gao2015}
Gao, W., Alcauter, S., Elton, A., Hernandez-Castillo, C.~R., Smith, J.~K.,
  Ramirez, J., and Lin, W. (2015).
\newblock {Functional network development during the first year: Relative
  sequence and socioeconomic correlations}.
\newblock {\em Cerebral Cortex}, 25(9):2919--2928.

\bibitem[Gonzalez-Castillo and Bandettini, 2018]{Gonzalez-Castillo2018}
Gonzalez-Castillo, J. and Bandettini, P.~A. (2018).
\newblock {Task-based dynamic functional connectivity: Recent findings and open
  questions}.
\newblock {\em NeuroImage}, 180(August 2017):526--533.

\bibitem[Hastings, 1970]{Hastings1970}
Hastings, W. (1970).
\newblock {Monte Carlo sampling methods using Markov chains and their
  applications}.
\newblock {\em Biometrika}, 57(1):97--109.

\bibitem[Heo et~al., 2022]{Heo2022}
Heo, K.~S., Shin, D.~H., Hung, S.~C., Lin, W., Zhang, H., Shen, D., and Kam,
  T.~E. (2022).
\newblock {Deep attentive spatio-temporal feature learning for automatic
  resting-state fMRI denoising}.
\newblock {\em NeuroImage}, 254(May 2021):119127.

\bibitem[Hoffmann et~al., 2020]{Hoffmann2020}
Hoffmann, T., Peel, L., Lambiotte, R., and Jones, N.~S. (2020).
\newblock {Community detection in networks without observing edges}.
\newblock {\em Science Advances}, 6(4):1--12.

\bibitem[Howell et~al., 2019]{Howell2019}
Howell, B.~R., Styner, M.~A., Gao, W., Yap, P.~T., Wang, L., Baluyot, K.,
  Yacoub, E., Chen, G., Potts, T., Salzwedel, A., Li, G., Gilmore, J.~H.,
  Piven, J., Smith, J.~K., Shen, D., Ugurbil, K., Zhu, H., Lin, W., and Elison,
  J.~T. (2019).
\newblock {The UNC/UMN Baby Connectome Project (BCP): An overview of the study
  design and protocol development}.
\newblock {\em NeuroImage}, 185(March 2018):891--905.

\bibitem[Hutchison et~al., 2013]{Hutchison2013}
Hutchison, R.~M., Womelsdorf, T., Allen, E.~A., Bandettini, P.~A., Calhoun,
  V.~D., Corbetta, M., Penna, S.~D., Duyn, J.~H., Glover, G.~H.,
  Gonzalez-castillo, J., Handwerker, D.~A., Keilholz, S., Kiviniemi, V.,
  Leopold, D.~A., Pasquale, F.~D., Sporns, O., Walter, M., and Chang, C.
  (2013).
\newblock {Dynamic functional connectivity : Promise , issues , and
  interpretations}.
\newblock {\em NeuroImage}, 80:360--378.

\bibitem[Kringelbach and Deco, 2020]{Kringelbach2020}
Kringelbach, M.~L. and Deco, G. (2020).
\newblock {Brain states and transitions: Insights from computational
  neuroscience}.
\newblock {\em Cell Reports}, 32(10):108128.

\bibitem[Lambiotte et~al., 2008]{Lambiotte2008}
Lambiotte, R., Delvenne, J.~C., and Barahona, M. (2008).
\newblock {Laplacian Dynamics and Multiscale Modular Structure in Networks}.
\newblock pages 1--29.

\bibitem[Lehmann et~al., 2021]{Lehmann2021}
Lehmann, B.~C., Henson, R.~N., Geerligs, L., Cam-CAN, and White, S.~R. (2021).
\newblock {Characterising group-level brain connectivity: A framework using
  Bayesian exponential random graph models}.
\newblock {\em NeuroImage}, 225:117480.

\bibitem[Lurie et~al., 2020]{Lurie2020}
Lurie, D.~J., Kessler, D., Bassett, D.~S., Betzel, R.~F., Breakspear, M.,
  Kheilholz, S., Kucyi, A., Li{\'{e}}geois, R., Lindquist, M.~A., McIntosh,
  A.~R., Poldrack, R.~A., Shine, J.~M., Thompson, W.~H., Bielczyk, N.~Z., Douw,
  L., Kraft, D., Miller, R.~L., Muthuraman, M., Pasquini, L., Razi, A.,
  Vidaurre, D., Xie, H., and Calhoun, V.~D. (2020).
\newblock {Questions and controversies in the study of time-varying functional
  connectivity in resting fMRI}.
\newblock {\em Network Neuroscience}, 4(1):30--69.

\bibitem[Metropolis et~al., 1953]{Metropolis1953}
Metropolis, N., Rosenbluth, A.~W., Rosenbluth, M.~N., Teller, A.~H., and
  Teller, E. (1953).
\newblock {Equation of state calculations by fast computing machines}.
\newblock {\em The Journal of Chemical Physics}, 21(6):1087--1092.

\bibitem[Monti et~al., 2017]{Monti2017a}
Monti, R.~P., Lorenz, R., Braga, R.~M., Anagnostopoulos, C., Leech, R., and
  Montana, G. (2017).
\newblock {Real-time estimation of dynamic functional connectivity networks}.
\newblock {\em Human Brain Mapping}, 38(1):202--220.

\bibitem[Newman, 2004]{Newman2004}
Newman, M.~E. (2004).
\newblock {Fast algorithm for detecting community structure in networks}.
\newblock {\em Physical Review E - Statistical Physics, Plasmas, Fluids, and
  Related Interdisciplinary Topics}, 69(6):5.

\bibitem[Newman, 2006a]{Newman2006a}
Newman, M.~E. (2006a).
\newblock {Modularity and community structure in networks}.
\newblock {\em Proceedings of the National Academy of Sciences of the United
  States of America}, 103(23):8577--8582.

\bibitem[Newman, 2006b]{Newman2006}
Newman, M. E.~J. (2006b).
\newblock {Modularity and community structure in networks}.
\newblock {\em PNAS}, 103(23):8577--8582.

\bibitem[Nobile and Fearnside, 2007]{Nobile2007}
Nobile, A. and Fearnside, A.~T. (2007).
\newblock {Bayesian finite mixtures with an unknown number of components: The
  allocation sampler}.
\newblock {\em Statistics and Computing}, 17:147--162.

\bibitem[Ogawa et~al., 1990]{Ogawa1990}
Ogawa, S., Lee, T.~M., Kay, A.~R., and Tank, D.~W. (1990).
\newblock {Brain magnetic resonance imaging with contrast dependent on blood
  oxygenation}.
\newblock {\em Proceedings of the National Academy of Sciences of the United
  States of America}, 87(24):9868--9872.

\bibitem[Park and Friston, 2013]{Park2013}
Park, H.-J. and Friston, K. (2013).
\newblock {Structural and functional brain networks: From connections to
  cognition}.
\newblock {\em Science}, 342(6158).

\bibitem[Pavlovi{\'{c}} et~al., 2020]{Pavlovic2020}
Pavlovi{\'{c}}, D.~M., Guillaume, B.~R., Towlson, E.~K., Kuek, N.~M., Afyouni,
  S., V{\'{e}}rtes, P.~E., {Thomas Yeo}, B., Bullmore, E.~T., and Nichols,
  T.~E. (2020).
\newblock {Multi-subject Stochastic Blockmodels for adaptive analysis of
  individual differences in human brain network cluster structure}.
\newblock {\em NeuroImage}, page 116611.

\bibitem[Peel et~al., 2017]{Peel2017}
Peel, L., Larremore, D.~B., and Clauset, A. (2017).
\newblock {The ground truth about metadata and community detection in
  networks}.
\newblock {\em Science Advances}, 3(5):1--9.

\bibitem[Razi and Friston, 2016]{Razi2016}
Razi, A. and Friston, K.~J. (2016).
\newblock {The connected brain: Causality, models, and intrinsic dynamics}.
\newblock {\em IEEE Signal Processing Magazine}, 26(5):340--343.

\bibitem[Razi et~al., 2015]{Razi2015}
Razi, A., Kahan, J., Rees, G., and Friston, K.~J. (2015).
\newblock {Construct validation of a DCM for resting state fMRI}.
\newblock {\em NeuroImage}, 106:1--14.

\bibitem[Razi et~al., 2017]{Razi2017}
Razi, A., Seghier, M.~L., Zhou, Y., McColgan, P., Zeidman, P., Park, H.-J.,
  Sporns, O., Rees, G., and Friston, K.~J. (2017).
\newblock {Large-scale DCMs for resting-state fMRI}.
\newblock {\em Network Neuroscience}, 1(4):381--414.

\bibitem[Richardson et~al., 2010]{Richardson2010}
Richardson, H.~L., Walker, A.~M., and Horne, R.~S. (2010).
\newblock {Sleeping like a baby - Does gender influence infant arousability?}
\newblock {\em Sleep}, 33(8):1055--1060.

\bibitem[Robinson et~al., 2015]{Robinson2015}
Robinson, L.~F., Atlas, L.~Y., and Wager, T.~D. (2015).
\newblock {Dynamic functional connectivity using state-based dynamic community
  structure: Method and application to opioid analgesia}.
\newblock {\em NeuroImage}, 108:274--291.

\bibitem[Rubinov and Sporns, 2010]{Rubinov2010}
Rubinov, M. and Sporns, O. (2010).
\newblock {Complex network measures of brain connectivity: uses and
  interpretations.}
\newblock {\em NeuroImage}, 52(3):1059--69.

\bibitem[Schaefer et~al., 2018]{Schaefer2018}
Schaefer, A., Kong, R., Gordon, E.~M., Laumann, T.~O., Zuo, X.-N., Holmes,
  A.~J., Eickhoff, S.~B., and Yeo, B. T.~T. (2018).
\newblock {Local-Global Parcellation of the Human Cerebral Cortex from
  Intrinsic Functional Connectivity MRI}.
\newblock {\em Cerebral Cortex}, 28(9):3095--3114.

\bibitem[Taghia et~al., 2018]{Taghia2018a}
Taghia, J., Cai, W., Ryali, S., Kochalka, J., Nicholas, J., Chen, T., and
  Menon, V. (2018).
\newblock {Uncovering hidden brain state dynamics that regulate performance and
  decision-making during cognition}.
\newblock {\em Nature Communications}, 9(1).

\bibitem[Ting et~al., 2021]{Ting2021}
Ting, C.~M., Samdin, S.~B., Tang, M., and Ombao, H. (2021).
\newblock {Detecting dynamic community structure in functional brain networks
  across individuals: A multilayer approach}.
\newblock {\em IEEE Transactions on Medical Imaging}, 40(2):468--480.

\bibitem[Vidaurre et~al., 2018]{Vidaurre2018}
Vidaurre, D., Abeysuriya, R., Becker, R., Quinn, A.~J., Alfaro-Almagro, F.,
  Smith, S.~M., and Woolrich, M.~W. (2018).
\newblock {Discovering dynamic brain networks from big data in rest and task}.
\newblock {\em NeuroImage}, 180(June 2017):646--656.

\bibitem[{Vos de Wael} et~al., 2020]{VosdeWael2020}
{Vos de Wael}, R., Benkarim, O., Paquola, C., Lariviere, S., Royer, J.,
  Tavakol, S., Xu, T., Hong, S.~J., Langs, G., Valk, S., Misic, B., Milham, M.,
  Margulies, D., Smallwood, J., and Bernhardt, B.~C. (2020).
\newblock {BrainSpace: a toolbox for the analysis of macroscale gradients in
  neuroimaging and connectomics datasets}.
\newblock {\em Communications Biology}, 3(1).

\bibitem[Wen et~al., 2019]{wen2019}
Wen, X., Zhang, H., Li, G., Liu, M., Yin, W., Lin, W., Zhang, J., and Shen, D.
  (2019).
\newblock {First-year development of modules and hubs in infant brain
  functional networks}.
\newblock {\em NeuroImage}, 185(October 2018):222--235.

\bibitem[Wyse and Friel, 2012]{Wyse2012}
Wyse, J. and Friel, N. (2012).
\newblock {Block clustering with collapsed latent block models}.
\newblock {\em Statistics and Computing}, 22:415--428.

\bibitem[Zhang et~al., 2019]{Zhang2019}
Zhang, H., Shen, D., and Lin, W. (2019).
\newblock {Resting-state functional MRI studies on infant brains: A decade of
  gap-filling efforts}.
\newblock {\em NeuroImage}, 185(July 2018):664--684.

\end{thebibliography}
%
%\bibliographystyle{unsrt}
%\bibliographystyle{apalike}
%bibliographystyle{nejm}

\newpage

\section*{\Huge \textbf{Supplementary information:}}

\captionsetup[figure]{labelfont={bf},name={SI Fig.},labelsep=period}

\renewcommand{\thefigure}{1}
\begin{figure*}[!ht]
\centering
\includegraphics[width=0.9\linewidth]{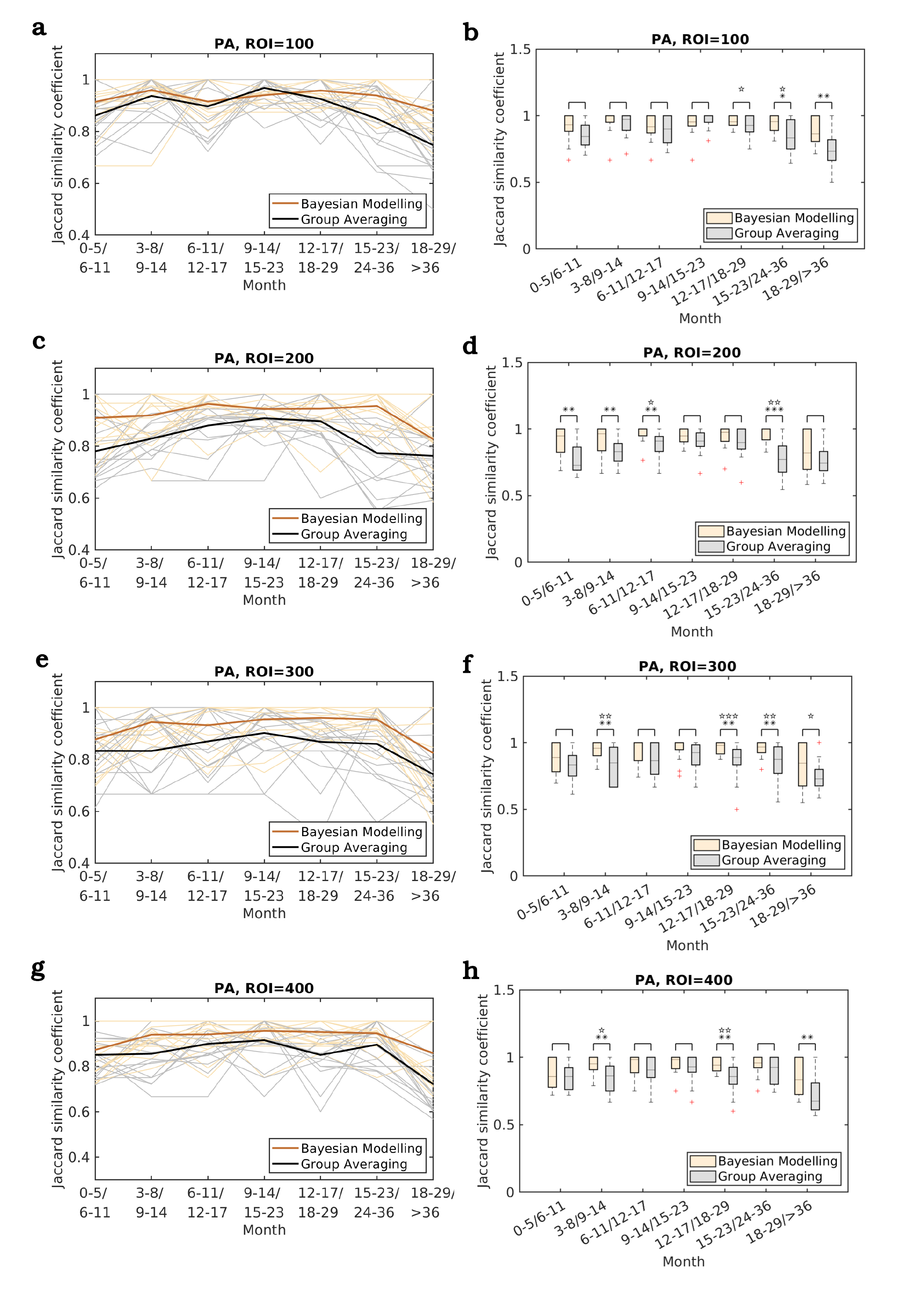}
\caption{\footnotesize \textbf{The comparison of the module evolution evaluated by our proposed Bayesian modelling method and conventional group averaging method (PA)}. The module evolution is evaluated for the infant brain networks with different numbers of ROIs (100 ROIs in \textbf{a} and \textbf{b}; 200 ROIs in \textbf{c} and \textbf{d}; 300 ROIs in \textbf{e} and \textbf{f}; and 400 ROIs in \textbf{g} and \textbf{h}). In the left panel, the light orange curves are the Jaccard similarity coefficients with different modularity resolutions $\gamma$ evaluated by Bayesian modelling method and grey lines are those evaluated by group averaging method. The darker orange curve and black curve are the mean values averaged across different values of $\gamma$. The boxplots in the right panel are the statistical analysis of the two methods. The star indicates the statistical significance of F-test (\ding{73}: $0.01<p<0.05$, \ding{73}\ding{73}: $0.001<p<0.01$, and \ding{73}\ding{73}\ding{73}: $p<0.001$) and the asterisk indicates the statistical significance of t-test ($\ast$: $0.01<p<0.05$, $\ast$ $\ast$: $0.001<p<0.01$, and $\ast$ $\ast$ $\ast$: $p<0.001$).}
\label{Jaccard_similarity_evolution}
\end{figure*}

\renewcommand{\thefigure}{2}
\begin{figure*}[!ht]
\centering
\includegraphics[width=0.9\linewidth]{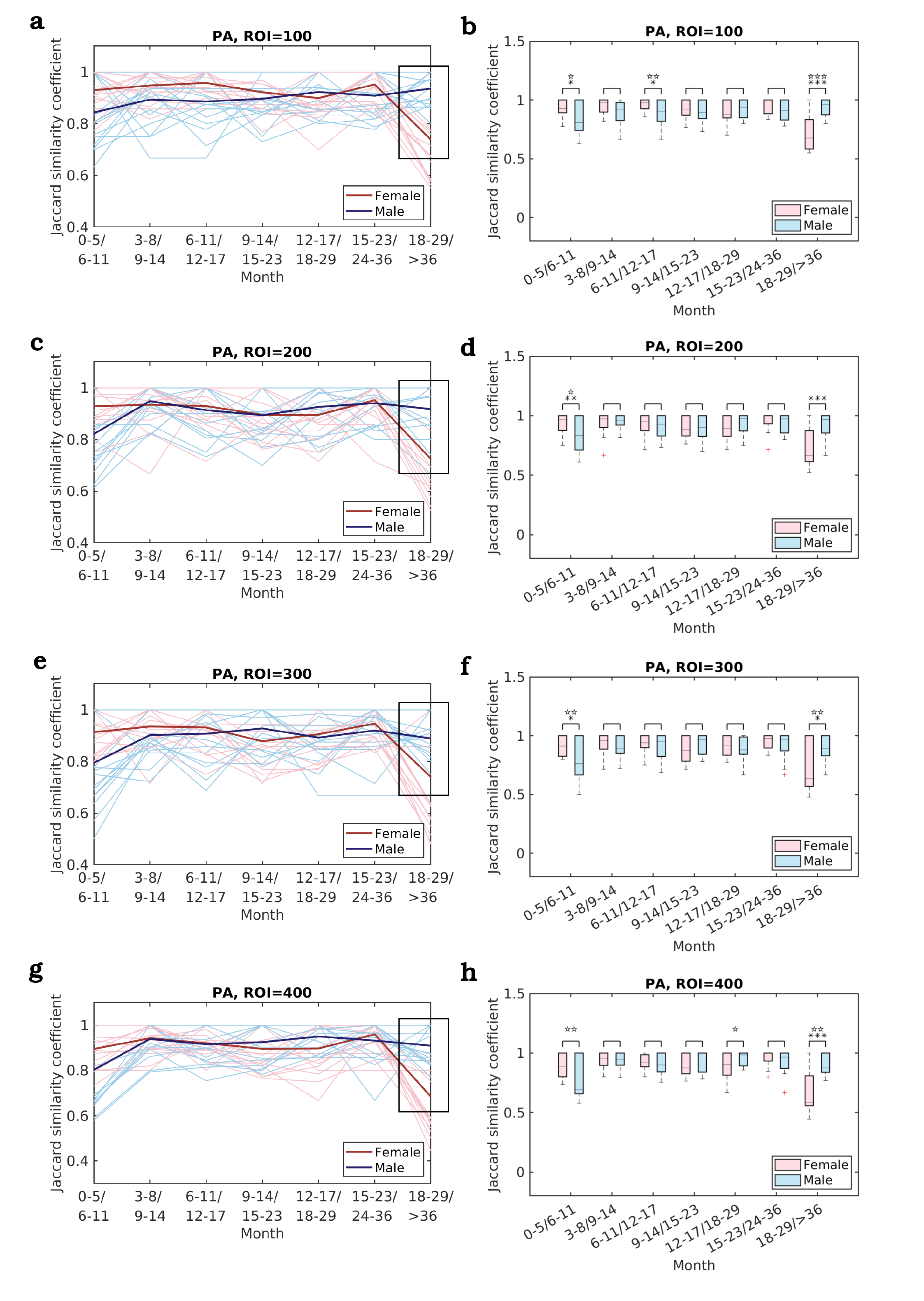}
\caption{\footnotesize \textbf{The comparison of the module evolution of female and male (PA)}. The module evolution is evaluated for the infant brain networks with different numbers of ROIs (100 ROIs in \textbf{a} and \textbf{b}; 200 ROIs in \textbf{c} and \textbf{d}; 300 ROIs in \textbf{e} and \textbf{f}; and 400 ROIs in \textbf{g} and \textbf{h}). In the left panel, the pink curves are the Jaccard similarity coefficients with different modularity resolutions $\gamma$ of female and light blue curves are those of the male. The darker pink curve and darker blue curve are the mean values averaged across different values of $\gamma$. The boxplots in the right panel are the statistical analysis of female and male. The star indicates the statistical significance of F-test (\ding{73}: $0.01<p<0.05$, \ding{73}\ding{73}: $0.001<p<0.01$, and \ding{73}\ding{73}\ding{73}: $p<0.001$) and the asterisk indicates the statistical significance of t-test ($\ast$: $0.01<p<0.05$, $\ast$ $\ast$: $0.001<p<0.01$, and $\ast$ $\ast$ $\ast$: $p<0.001$).}
\label{Jaccard_similarity_gender}
\end{figure*}

\renewcommand{\thefigure}{3}
\begin{figure*}[!ht]
\centering
\includegraphics[width=0.9\linewidth]{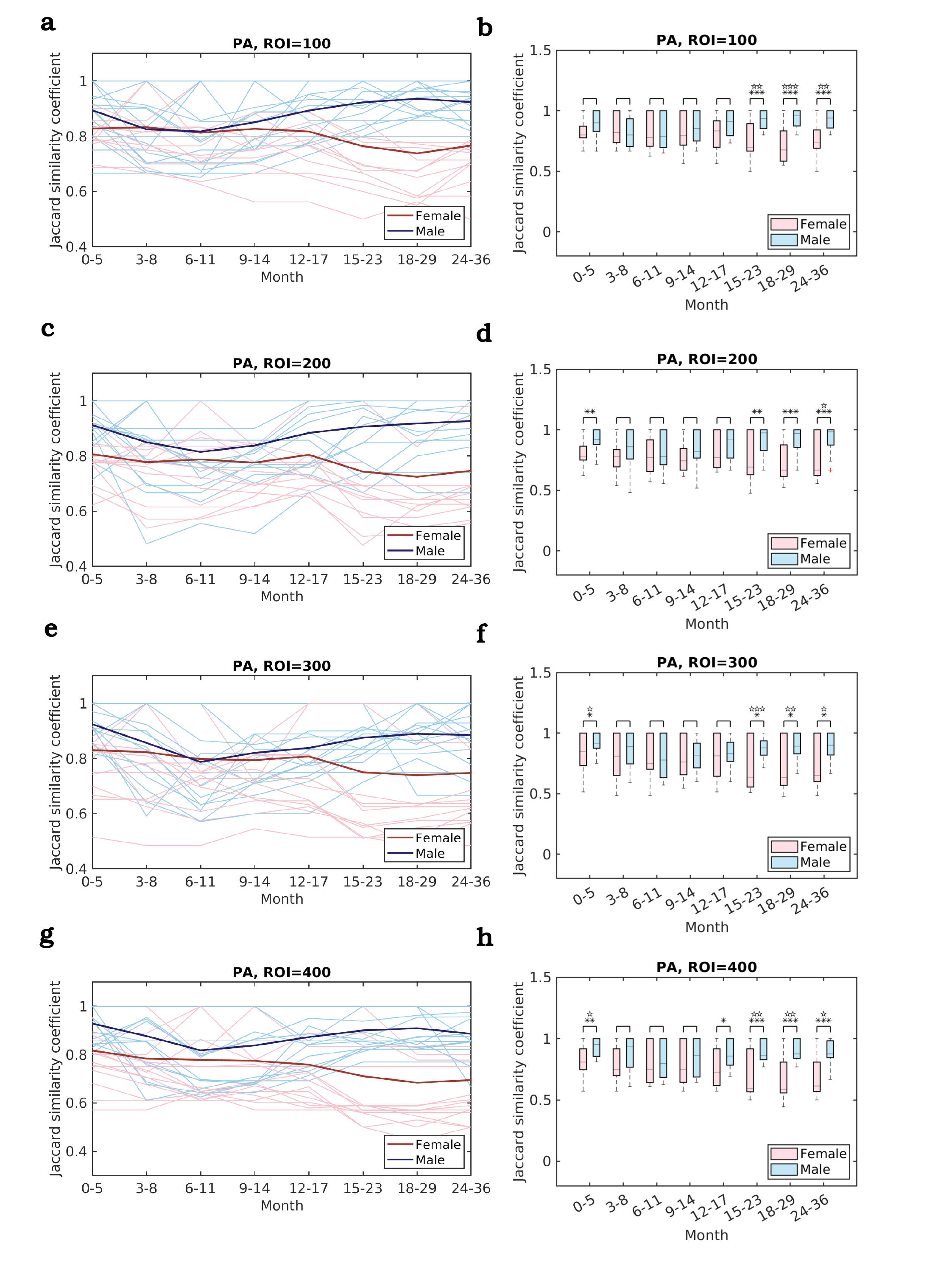}
\caption{\footnotesize \textbf{The similarity of modular structure between sleeping and awake conditions for female and male (PA)}. The Jaccard similarity between sleeping and awake conditions are evaluated for the infant brain networks with different numbers of ROIs (100 ROIs in \textbf{a} and \textbf{b}; 200 ROIs in \textbf{c} and \textbf{d}; 300 ROIs in \textbf{e} and \textbf{f}; and 400 ROIs in \textbf{g} and \textbf{h}). In the left panel, the pink curves are the Jaccard similarity coefficients with different modularity resolutions $\gamma$ of female and light blue curves are those of the male. The darker pink curve and darker blue curve are the mean values averaged across different values of $\gamma$. The boxplots in the right panel are the statistical analysis of female and male. The star indicates the statistical significance of F-test (\ding{73}: $0.01<p<0.05$, \ding{73}\ding{73}: $0.001<p<0.01$, and \ding{73}\ding{73}\ding{73}: $p<0.001$) and the asterisk indicates the statistical significance of t-test ($\ast$: $0.01<p<0.05$, $\ast$ $\ast$: $0.001<p<0.01$, and $\ast$ $\ast$ $\ast$: $p<0.001$). Note that the x-axis represents the age ranges of only sleeping condition, which is different from those in SI Fig.\ref{Jaccard_similarity_gender}.}
\label{Jaccard_similarity_awake_sleep}
\end{figure*}

\end{document}